\documentclass[aps,11pt,twocolumn,superscriptaddress,balancelastpage,showpacs,reprint]{revtex4-1}
\usepackage{graphicx}
\usepackage[caption=false]{subfig}
\usepackage{dcolumn}
\usepackage{bm}
\usepackage{mathtools}
\usepackage{color,soul}
\usepackage{multirow}
\usepackage{amsmath}
\usepackage{tikz}

\usepackage[section]{placeins}
\usepackage{titlesec}

\titlespacing\section{0pt}{10pt plus 4pt minus 2pt}{0pt plus 2pt minus 2pt}
\titlespacing\subsection{0pt}{10pt plus 4pt minus 2pt}{0pt plus 2pt minus 2pt}
\titlespacing\subsubsection{0pt}{10pt plus 4pt minus 2pt}{0pt plus 2pt minus 2pt}

\begin{document}
\title{Spatial gene drives and pushed genetic waves}

\author{Hidenori Tanaka}
\email{tanaka@g.harvard.edu}
\affiliation{School of Engineering and Applied Sciences, Harvard University, Cambridge, MA 02138}
\affiliation{Kavli Institute for Bionano Science and Technology, Harvard University, Cambridge, MA 02138}

\author{Howard A. Stone}
\affiliation{Department of Mechanical and Aerospace Engineering, Princeton University, NJ 08544, USA}

\author{David R. Nelson}
\email{drnelson@fas.harvard.edu}
\affiliation{School of Engineering and Applied Sciences, Harvard University, Cambridge, MA 02138}
\affiliation{Departments of Physics and Molecular and Cellular Biology, Harvard University, Cambridge, MA 02138, USA}

\date{\today}
\begin{abstract}
Gene drives have the potential to rapidly replace a harmful wild-type allele with a gene drive allele engineered to have desired functionalities. However, an accidental or premature release of a gene drive construct to the natural environment could damage an ecosystem irreversibly. Thus, it is important to understand the spatiotemporal consequences of the super-Mendelian population genetics prior to potential applications. Here, we employ a reaction-diffusion model for sexually reproducing diploid organisms to study how a locally introduced gene drive allele spreads to replace the wild-type allele, even though it possesses a selective disadvantage $s>0$. Using methods developed by N. Barton and collaborators, we show that socially responsible gene drives require $0.5<s<0.697$, a rather narrow range. In this ``pushed wave'' regime, the spatial spreading of gene drives will be initiated only when the initial frequency distribution is above a threshold profile called ``critical propagule'', which acts as a safeguard against accidental release. We also study how the spatial spread of the pushed wave can be stopped by making gene drives uniquely vulnerable (``sensitizing drive'') in a way that is harmless for a wild-type allele. Finally, we show that appropriately sensitized drives in two dimensions can be stopped even by imperfect barriers perforated by a series of gaps. 
\end{abstract}
\maketitle

The development of the CRISPR/Cas9 system \cite{cong2013multiplex, jinek2013rna,mali2013rna,wright2016biology}, derived from an adaptive immune system in prokaryotes \cite{marraffini2015crispr}, has received much recent attention, in part due to its exceptional versatility as a gene editor in sexually-reproducing organisms, compared to similar exploitations of homologous recombination such as zinc-finger nucleases (ZFNs) and the TALENS system \cite{jiang2015crispr,wright2016biology}. Part of the appeal is the potential for introducing a novel gene into a population, allowing control of highly pesticide-resistant crop pests and disease vectors such as mosquitoes \cite{alphey2014genetic, burt2014heritable, esvelt2014concerning,gantz2016dawn}. Although the genetic modifications typically introduce a fitness cost or a ``selective disadvantage'', 
the enhanced inheritance rate embodied in CRISPR/Cas9 gene drives nevertheless allows edited genes to spread, even when the fitness cost of the inserted gene is large. 
The idea of using constructs that bias gene transmission rates to rapidly introduce novel genes into ecosystems has been discussed for many decades \cite{curtis1968possible, foster1972chromosome, burt2003site, sinkins2006gene, gould2008broadening, deredec2008population}. Similar ``homing endonuclease genes'' (in the case of CRISPR/Cas9, the homing ability is provided by a guide RNA) were considered earlier by ecologists in the context of control of malaria in Africa \cite{north2013modelling,Eckhoff:2017:10.1073/pnas.1611064114}.

\begin{figure}
\centering
\includegraphics[clip,width=0.9\columnwidth]{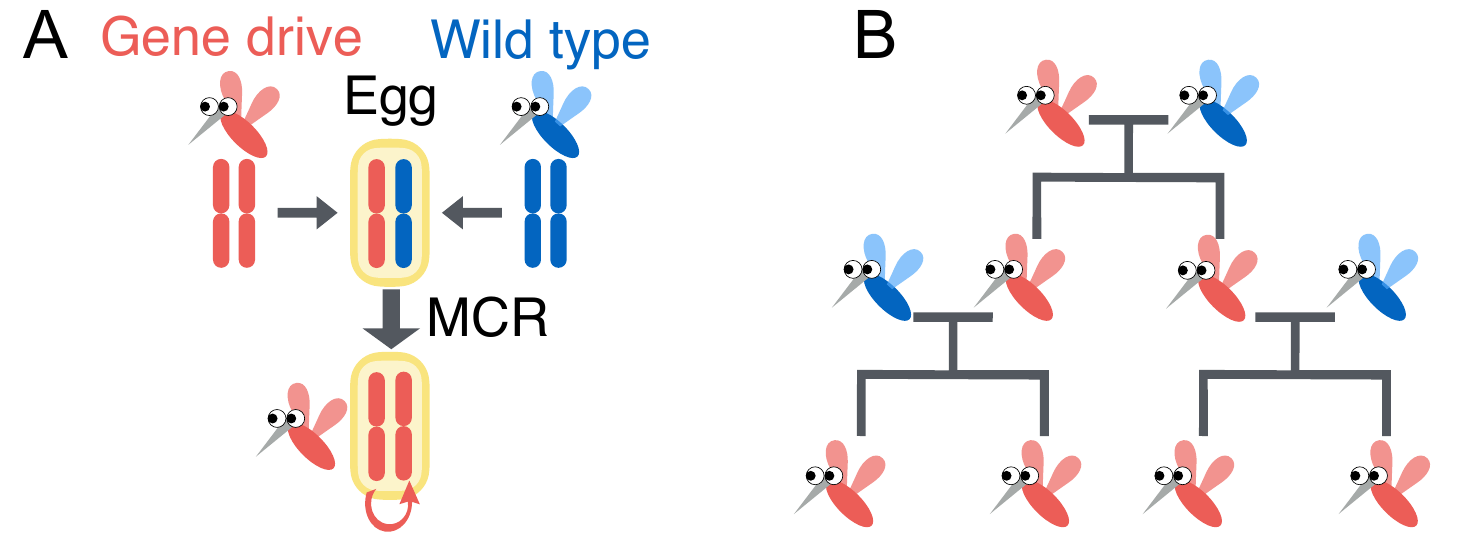}
\caption{
Schematics of the gene drive machinery with a perfect conversion efficiency $c=1$. 
(A) Every time an individual homozygous for the drive construct and a wild-type mate, heterozygotes in the embryo are converted to homozygotes by the mutagenic chain reaction (MCR).
(B) Gene drives enhance their inheritance rate beyond that of the conventional Mendelian population genetics and can spread even with a selective disadvantage. 
}
\label{Fig_1}
\end{figure} 

As a hypothetical example of a gene drive applied to a pathogen vector requiring both a vertebrate and insect host, consider plasmodium, carried by mosquitoes and injected with its saliva into humans (Fig. \ref{Fig_1}). Female mosquitoes typically hatch from eggs in small standing pools of water and, after mating, search for a human to feed on. They then lay their eggs and repeat the process, thus spreading the infection over a few gonotrophic cycles. A gene drive could alter the function of a protein manufactured in the salivary gland of female mosquitoes from, say, type $a$, anesthetizing nerve cells when it bites humans, to instead type $A$, clogging up essential chemoreceptors in plasmodium and thus killing these eukaryotes. In the absence of a gene drive, there would be a selective disadvantage or fitness cost $s$ to losing this protein. Even if the fitness cost $s$ were zero, it is unlikely that this new trait would be able to escape genetic drift in large populations. However, as we describe below, the trait could spread easily if linked to a gene drive that converts heterozygotes to homozygotes with efficiency $c$ close to 1 (Fig. \ref{Fig_1}A). 
Remarkably, high conversion rates have already been achieved with the mutagenic chain reaction (``MCR'') realized by the CRISPR/Cas9 system \cite{cong2013multiplex, jinek2013rna,mali2013rna} for yeast ($c_{\rm{yeast}}>0.995$) \cite{dicarlo2015safeguarding}, fruit flies ($c_{\rm{flies}}=0.97$) \cite{gantz2015mutagenic} and malaria vector mosquito, \emph{Anopheles stephensi} with engineered malaria resistance ($c_{\rm{mosquito}}\geq 0.98$) \cite{gantz2015highly}.

However, the gene drives' intrinsic nature of irreversibly altering wild-type populations raises biosafety concerns \cite{esvelt2014concerning}, and calls for confinement strategies to prevent unintentional escape and spread of the gene drive constructs \cite{akbari2015safeguarding}.
While various genetic design or containment strategies have been discussed \cite{chan2011insect, henkel2012monitoring, esvelt2014concerning, gantz2015mutagenic}, and a few computational simulations were conducted \cite{huang2011gene,north2013modelling,Eckhoff:2017:10.1073/pnas.1611064114}, the \emph{spatial} spreading of the gene drive alleles has received less attention. 

To understand such phenomena in a spatial context, we will exploit a methodology developed by N. Barton and collaborators, originally in an effort to understand adaptation and speciation of diploid sexually reproducing organisms in genetic hybrid zones \cite{barton1979dynamics, barton1989adaptation, barton2011spatial}. We apply these techniques to a spatial generalization of a model of diploid CRISPR/Cas9 population genetics proposed by Unckless \emph{et al.} \cite{unckless2015modeling}, and highlight two distinct ways in which gene drive alleles can spread spatially. The non-Mendelian (or ``super-Mendelian'' \cite{Noble057281}) population genetics of gene drives are remarkable because individuals homozygous for a gene drive can in fact spread into wild-type populations even if they carry a positive selective disadvantage $s$ (Fig. \ref{Fig_1}B).   First, for small selective disadvantages ($0 < s < 0.5$ in our case), the spatial spreading proceeds via a well-known Fisher-Kolmogorov-Petrovsky-Piskunov wave \cite{fisher1937wave, kolmogorov}. Such pulled genetic waves \cite{stokes1976two,lewis2016finding,gandhi2016range} are driven by growth and diffusive dispersal at the leading edge, and are difficult to slow down and stop.   

However, for somewhat larger selective disadvantages ($0.5 < s < 0.697$) we find that propagation proceeds instead via a pushed genetic wave \cite{stokes1976two,lewis2016finding,gandhi2016range}, where the genetic wave advances via accentuated growth from populations somewhat behind the front that spill over the leading edge. These waves, characterized by a strong Allee effect \cite{lewis1993allee, taylor2005allee}, are more socially responsible than the pulled Fisher waves because: (i) only inoculations whose spatial size and density exceed a critical nucleus, or ``critical propagule''\cite{barton2011spatial} are able to spread spatially, thus providing protection against a premature or accidental release of a gene drive, (ii) the gene drive pushed waves can be stopped by making them uniquely vulnerable to a specific compound (``sensitizing drive'' \cite{esvelt2014concerning}), which is harmless for a wild-type allele, and (iii) appropriately sensitized gene drives can be stopped even by barriers punctuated by defects, analogous to regularly spaced fire breaks used to contain forest fires. 
Similar pushed or ``excitable'' waves also arise, for example, in neuroscience, in simplified versions of the Hodgkin-Huxley model of action potentials \cite{nelson2004biological}. When the selective disadvantage associated with the gene drive is too large ($s > 0.697$ in our model) the excitable wave reverses direction and the region occupied by the gene drive homozygotes collapses to zero.

\begin{figure}
\centering
\includegraphics[clip,width=0.8\columnwidth]{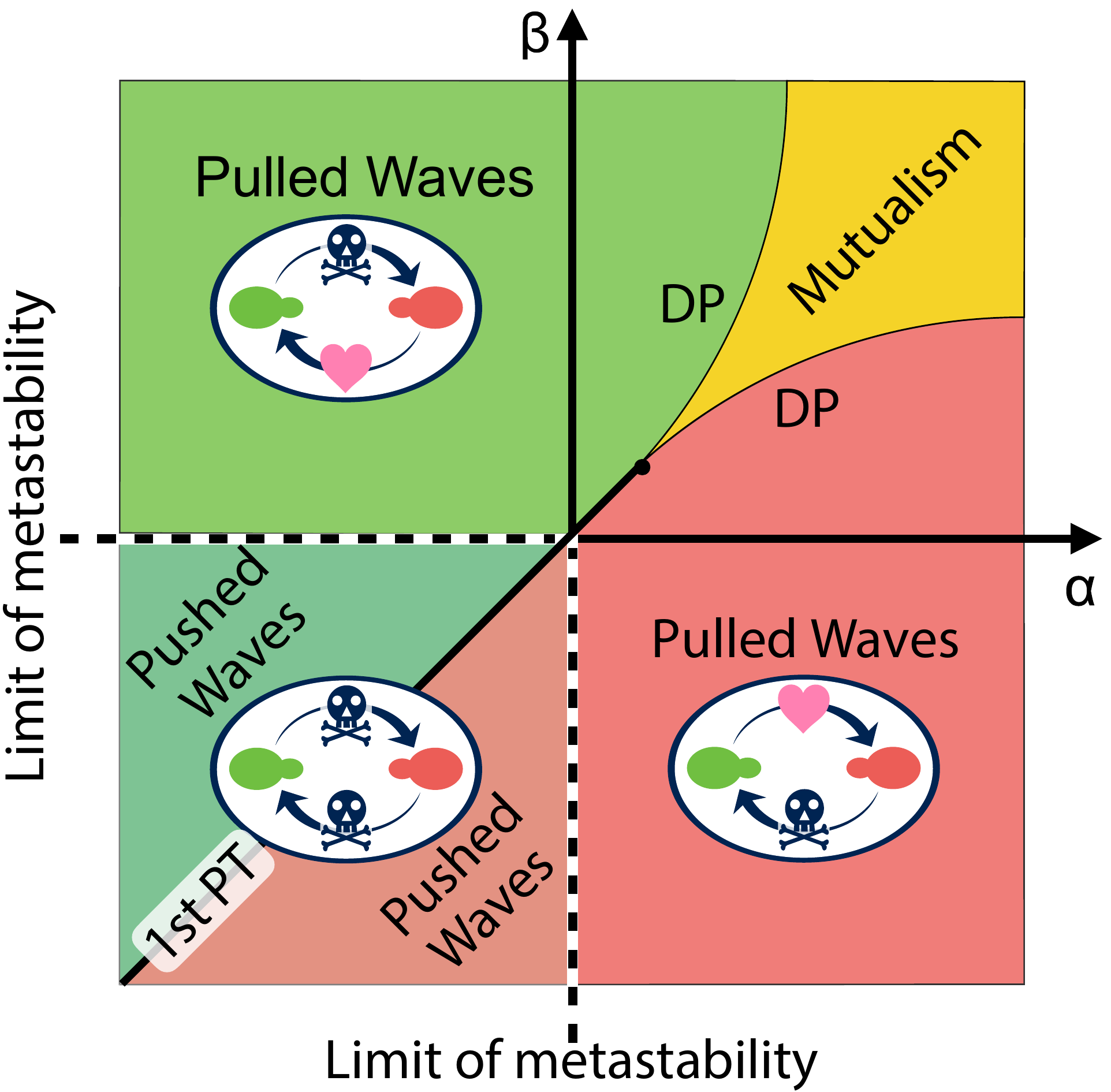}
\caption{
Schematic phase diagram of the spatial evolutionary games in one dimension \cite{frey2010evolutionary, korolev2011competition,lavrentovich2014asymmetric}.
The parameters $\alpha$ and $\beta$ control interactions between red and green haploid organisms. Positive $\alpha$ means the presence of the green allele favors the red allele, positive $\beta$ enhances the green allele when red is present, etc. (see SI Appendix for a detailed description of the model.) Pulled Fisher wave regimes (controlling, for example, the dynamics of selective dominance in the second and four quadrants) and the pushed excitable wave regimes (third quadrant, competitive exclusion dynamics) are bounded by the black dashed spinodal lines $\alpha=0, \beta < 0$ and $\alpha<0, \beta = 0$. These two bistable regimes are separated by  the first-order phase transition (PT) line $\alpha=\beta<0$, drawn as a black solid line.
}
\label{Fig_2}
\end{figure} 

The same mathematical analyses applies to spatial evolutionary games of two competing species in one dimension, which are governed by a class of reaction-diffusion equations that resemble the gene drive system. The fitnesses of the two interacting red and green species ($w_R$, $w_G$) are related to their frequencies ($f(x,t)$, $1-f(x,t)$) by $w_R (x,t) = g + \alpha (1-f(x,t)), w_G (x,t) = g + \beta f(x,t)$, where $g$ is a background fitness, assumed identical for the two alleles for simplicity.
The mutualistic regime $\alpha>0, \beta>0$ in the first quadrant of Fig.~\ref{Fig_2} has been studied already \cite{korolev2011competition}, including the effect of genetic drift, with two lines of directed ``percolation'' transitions out of a mutualistic phase. 
Here, we apply the methods of \cite{barton2011spatial} to study the evolutionary dynamics near the line of first-order transitions that characterize the competitive exclusion regime in the third quadrant of Fig.~\ref{Fig_2}.
Because the mathematics parallels the analysis inspired by gene drive systems in the main text, we relegate discussion of this topic to the SI Appendix, which also discusses conversion efficiencies $c<1$, an analogy with nucleation theory, laboratory tests and other matters.\\

\subsection*{Mathematical model of the CRISPR gene drives}
We start with a Hardy-Weinberg model \cite{hartl1997principles} and incorporate a mutagenic chain reaction (``MCR'') with $100\%$ conversion rate to construct a model for a well-mixed system.
This model is the limiting case of ``$c=1$'' in the work of Unckless \emph{et al} \cite{unckless2015modeling}.
Conversion efficiencies $c<1$ can be handled by similar techniques. 
First, we consider a well-mixed diploid system with a wild-type allele $a$ and a gene drive allele $A$ with frequencies $p=p(t)$ and $q=q(t)$ respectively at time $t$, with $p(t)+q(t)=1$.
Within a random mating model, the allele frequencies after one generation time $\tau_g$ are given by
\begin{equation}
(pa+qA)^2 = p^2 (a,a) + 2pq(a,A)+q^2(A,A),
\end{equation}
and the ratios of fertilized eggs with diploid types $(a,a)$, $(a,A)$ and $(A,A)$ are $p^2:2pq:q^2$.
In a heterozygous $(a,A)$ egg, the CRISPR/Cas9 machinery encoded on a gene drive allele $A$ converts the wild-type allele $a$ into a gene drive allele $A$. 
Here, we assume a perfect conversion rate $(a,A)\xrightarrow[\rm{MCR}]{c=1}(A,A)$ in the embryo, as has been approximated already for yeast \cite{dicarlo2015safeguarding} and fruit flies \cite{gantz2015mutagenic}. 
Genetic engineering will typically reduce the fitness of individuals carrying the gene drive alleles compared to wild-type organisms, which have already gone through natural evolution and may be near a fitness maximum.

The selective disadvantage of a gene drive allele $s$ is defined by the ratio of the fitness $w_{\rm{wild}}$ of wild-type organisms $(a,a)$ to the fitness $w_{\rm{drive}}$ of $(A,A)$ individuals carrying the gene drive,
\begin{equation}
\frac{w_{\rm{drive}}}{w_{\rm{wild}}}\equiv1-s,~0\leq s.
\end{equation}
(In the limit $c\rightarrow 1$ no heterozygous $(a,A)$ individuals are born \cite{unckless2015modeling}.)
Taking the fitness into account, the allele frequencies after one generation time $\tau_g$ are
\begin{equation}
p' : q' = w_{\rm{wild}} p^2 : w_{\rm{wild}} (1-s)(q^2 + 2pq),
\end{equation}
where $p'\equiv p(t+\tau_g)$ and $q'\equiv q(t+\tau_g)$.
Upon approximating $q'-q=q(t+\tau_g)-q(t)$ by $\tau_g \frac{dq}{dt}$, we obtain a differential equation
\begin{equation}
\begin{split} 
\tau_g \frac{dq}{dt} &= \frac{(1-s)(q^2 + 2pq)}{p^2+(1-s)(q^2 + 2pq)} - q\\
&=\frac{sq(1-q)(q-q^*)}{1-sq(2-q)},\textnormal{ where}~q^* = \frac{2s-1}{s},
\end{split}
\label{eq4}
\end{equation}
which governs population dynamics of the mutagenic chain reaction with $100\%$ conversion efficiency in a well-mixed system. 
To take spatial dynamics into account, we add a diffusion term \cite{barton2011spatial} and obtain a deterministic reaction-diffusion equation for the MCR model, namely
\begin{equation} \label{rdMCR}
\tau_g \frac{\partial q}{\partial t} = \tau_g D \frac{\partial^2 q}{\partial x^2}+ \frac{sq(1-q)(q-q^*)}{1-sq(2-q)},
\end{equation}
which will be the main focus of this article.
For later discussions, we name the reaction term of the reaction-diffusion equation, 
\begin{equation}
f_{\rm{MCR}}(q,s) = \frac{sq(1-q)(q-q^*)}{1-sq(2-q)}.
\label{FMCR}
\end{equation}
The reaction term reduces to a simpler cubic expression 
\begin{equation}
f_{\rm{cubic}}(q,s)=sq(1-q)(q-q^*)
\label{Fcubic}
\end{equation}
 by ignoring $-sq(2-q)$ in the denominator, which is a reasonable approximation if the selective disadvantage $s$ is small.
 This form of the reaction-diffusion equation has been well studied, as reviewed in \cite{barton2011spatial}.\\

 Although population genetics is often studied in the limit of small $s$, $s$ is in fact fairly large in the regime of pushed excitable waves of most interest to us here, $0.5 < s < 1.0$.   Hence, we will keep the denominator of the reaction term, as was also done in  \cite{barton2011spatial} with a different reaction term.   Comparison of results for the full nonlinear reaction term with those for the cubic approximation will give us a sense of the robustness of the cubic approximation.  Although it might also be of interest to study corrections to the continuous time approximation arising from higher order time derivatives in
 $(q'-q)/\tau_g = \frac{\partial q}{\partial t} + \frac{1}{2} \tau_g \frac{\partial^2 q}{\partial t^2}+...$
 (contributions from $\tau_g \frac{\partial^2 q}{\partial t^2}$ are formally of order $s^2$ ), this complicated problem will be neglected here; see, however, \cite{turellibartonpre} for a study of the robustness of the continuous time approximation, motivated by a model of dengue-suppressing Wolbachia in mosquitoes.

\subsection*{Initiation of the pushed waves} 
\begin{figure}[t!]
\centering
\includegraphics[clip,width=0.8\columnwidth]{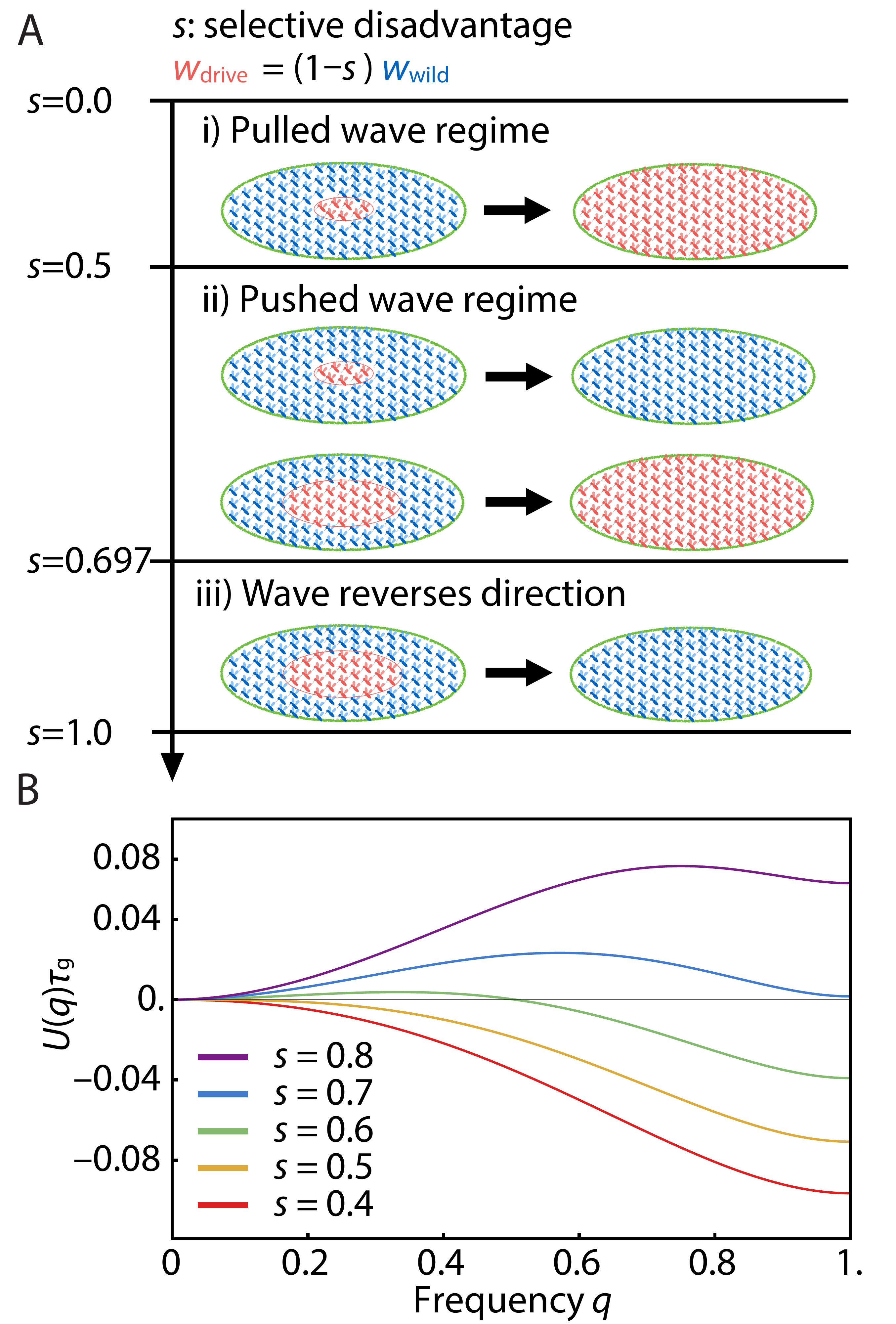}
\caption{
(A)~Spatial dynamics of gene drives can be determined by both the selective disadvantage $s$ and (when $0.5<s<0.697$), the size and intensity of the initial condition. (B) The energy landscapes $U(q)$ with various selective disadvantages $s$.
i) Pulled Fisher wave regime: When $s$ is small, $s\leq s_{\rm{min}}=0.5$ (lowermost red and yellow curves), fixation of the gene drive allele ($q=1$) is the unique stable state and there is no energy barrier between $q=0$ and $1$. Any finite introduction of a gene drive allele is sufficient to initiate a pulled Fisher population wave that spreads through space to saturate the system.
ii) Pushed excitable wave regime: When $s$ is slightly larger (green curve), and satisfies $s_{\rm{min}}=0.5<s<s_{\rm{max}}=0.697$, $q=1$ is still the preferred stable state, but an energy barrier at $q=q^*$ appears between $q=0$ and $1$. In this regime, the introduction of the gene drive allele at sufficient concentration and over a sufficiently large spatial extent is required for a pushed wave to spread to global fixation.
iii) Wave reverses direction: When $s$ is large, $s>s_{\rm{max}}= 0.697$ (topmost blue and purple curves), $q=0$ is the unique ground state and the gene drive species cannot establish a traveling population wave and so dies out.  
}
\label{Fig_3}
\end{figure} 

The reaction terms $f_{\rm{MCR}}(q,s)$ and $f_{\rm{cubic}}(q,s)$ have three identical fixed points,
$q=0,~1$ and $~q^*\big(=\frac{2s-1}{s}\big)$.
As discussed in the SI Appendix in connection to classical nucleation theory in physics, and following \cite{barton1979dynamics}, we can define the potential energy function 
\begin{equation}
U(q) = -\frac{1}{\tau_g} \int^{q}_{0} \frac{sq'(1-q')(q'-q^*)}{1-sq'(2-q')} dq'
\end{equation}
to identify qualitatively different parameter regimes.
In a well-mixed system, without spatial structure, the gene drive frequency $q(t)$ obeys Eq.~\ref{eq4}, and evolves in time so that it arrives at a local minimum of $U(q)$.
For the spatial model of interest here, $q(x,t)$ shows qualitatively distinct behaviors in three parameter regimes depending on the selective disadvantage $s$ (see Fig.~\ref{Fig_3}\emph{A}).
We plot the potential energy functions $U(q)$ in these parameter regimes in Fig.~\ref{Fig_3}\emph{B}.\\

i)~First, when $s<s_{\rm{min}}=0.5$, fixation of a gene drive allele $q(x)=1$ for all $x$ is the unique stable state and there is no energy barrier to reach the ground state starting from $q\approx0$.
In this regime, any finite frequency of gene drive allele locally introduced in space (provided it overcomes genetic drift) will spread and replace the wild-type allele.
The frequency profile will evolve as a pulled traveling wave $q(x,t)=Q(x-vt)$ with wave velocity $v$.
Such a wave was first found by Fisher \cite{fisher1937wave} and by Kolmogorov, Petrovsky and Piskunov \cite{kolmogorov} in the 1930s, in studies of how locally introduced organisms with advantageous genes spatially spread and replace inferior genes.
However, the threshold-less initiation of population waves of engineered gene drives with relatively small selective disadvantages seems highly undesirable, since the accidental escape of a single gene drive construct can establish a population wave that spreads freely into the extended environment.

ii)~There is a second regime for $0.5<s<0.697$ in which the potential energy function $U(q)$ exhibits an energy barrier between $q=0$ and $q=1$.
In this regime, a pushed traveling wave can be excited only when a threshold gene drive allele frequency is introduced over a sufficiently broad region of space that exceeds the size of a critical nucleus, which we investigate in the next section.
The existence of this threshold acts as a safeguard against accidental release. In addition, such excitable waves are easier to stop as we will discuss later. 
It appears that gene drives in this relatively narrow intermediate regime are the most desirable from a biosafety perspective.

iii)~When $s>s_{\rm{max}}=0.697$, the fixation of a gene drive allele throughout space is no longer absolutely stable (Fig.~\ref{Fig_3}\emph{B}), and a gene drive population wave cannot be established.
Indeed, the excitable wave reverses direction for $s>s_{\rm{max}}$.
An implicit equation for $s_{\rm{max}}$ results from equating $U(0)=U(1)=0$, which yields
\begin{equation}
\begin{split}
0 &= \int^{1}_{0} \frac{sq(1-q)(q-q^*)}{1-sq(2-q)}dq,\\
\textnormal{or } 0 &=\frac{-2+s_{\rm{max}}+2\sqrt{-1+\frac{1}{s_{\rm{max}}}}\arcsin(\sqrt{s_{\rm{max}}})}{2s_{\rm{max}}}\\
&\Rightarrow s_{\rm{max}}\approx 0.697,
\end{split}
\end{equation}
where we used $q^*=(2s-1)/s$. When $s>s_{\rm{max}}$, the locally introduced gene drive allele contracts rather than expands relative to the wild-type allele and simply dies out. See SI Appendix for the analogous results with an arbitrary conversion rate ($0<c<1$).

\subsection*{Critical nucleus in the pushed wave regime}
\begin{figure}[t!]
\centering
\includegraphics[clip,width=0.8\columnwidth]{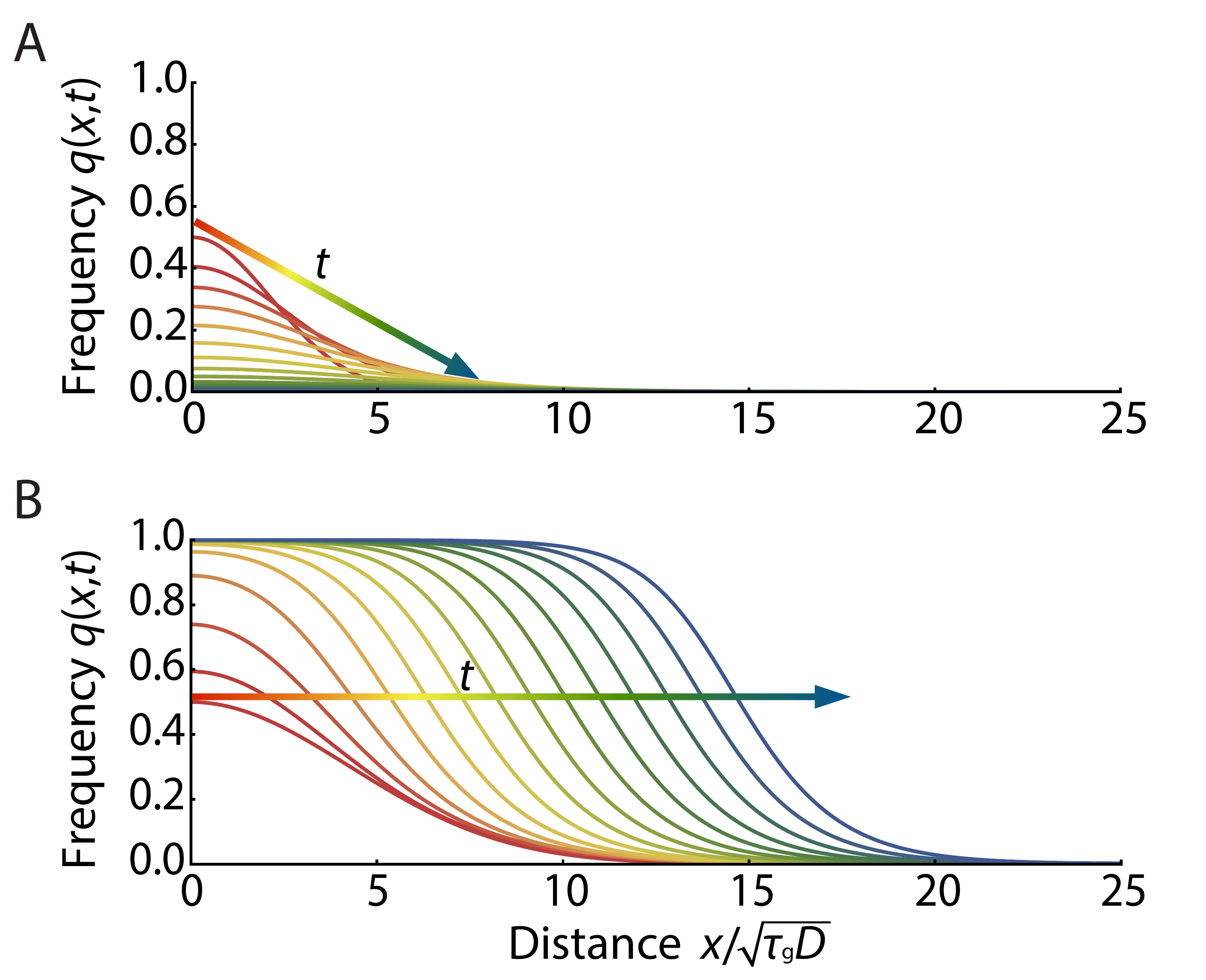}

\caption{
The excitable population wave carrying a gene drive can be established only when the initial concentration is above a threshold distribution and over a region of sufficient spatial extent (the critical nucleus or ``critical propagule'' \cite{barton2011spatial}).
Numerical solutions of
$\tau_g \frac{\partial q}{\partial t}=\tau_g D \frac{\partial^2 q}{\partial x^2} + \frac{sq(1-q)(q-q^*)}{1-sq(2-q)}$
 with $q^*=\frac{2s-1}{s}$ are plotted with time increment $\Delta t = 2.5 \tau_g$. 
The early time response is shown in red with later times in blue.
Selective disadvantage of the gene drive allele relative to the wild-type allele is set to $s=0.58$. 
In the case illustrated here, the gene drive allele can either die out or saturate the entire system, depending on the width of initial Gaussian population profile of $q(x,0)=ae^{-(x/B)^2}$.
(A) With a narrow distribution of the initially introduced gene drive species $(a=0.5, B=3.0\sqrt{\tau_g D})$, the population quickly fizzles out.
(B) With a broader distribution of the initial gene drive allele $(a=0.5, B=6.0\sqrt{\tau_g D})$, the gene drive allele successfully establishes a pushed population wave leading to $q(x)=1$ over the entire system.
}
\label{Fig_4}
\end{figure}
When the selective disadvantage $s$ is in the intermediate regime, $s_{\rm{min}}=1/2<s<s_{\rm{max}}= 0.697$, we can control initiation of the pushed excitable wave by the initial frequency profile of the gene drive allele $q(x,0)$ as shown in Fig.~\ref{Fig_4}.
For example, in Fig.~\ref{Fig_4}\emph{A}, an initially introduced gene drive allele (in the form of a Gaussian) diminishes and dies out since the width of the initial frequency distribution $q(x,0)$ is not sufficient to excite the population wave.
In contrast, the results in Fig.~\ref{Fig_4}\emph{B} show the successful establishment of the excitable wave starting from a sufficiently broad (Gaussian) initial distribution of a gene drive allele. 
Roughly speaking (provided $\frac{1}{2}<s<s_{\rm{max}}$), two conditions must be satisfied to obtain a critical propagule:
(1) The initial condition $q(0,0)$ at the center of the inoculant must exceed $q^* = \frac{2s-1}{s}$, the local maximum of the function $U(q)$ plotted in Fig.~\ref{Fig_3}; and 
(2) The spatial spread $\Delta x$ of the inoculant $q(x,t=0)$ must satisfy $\Delta x \gtrsim \rm{const}\sqrt{D \tau_g}$ where the dimensionless constant depends on $s$. Thus, the initial width should exceed the width of the pushed wave that is being launched.

\begin{figure}[b!]
\centering
\includegraphics[width=0.8\columnwidth]{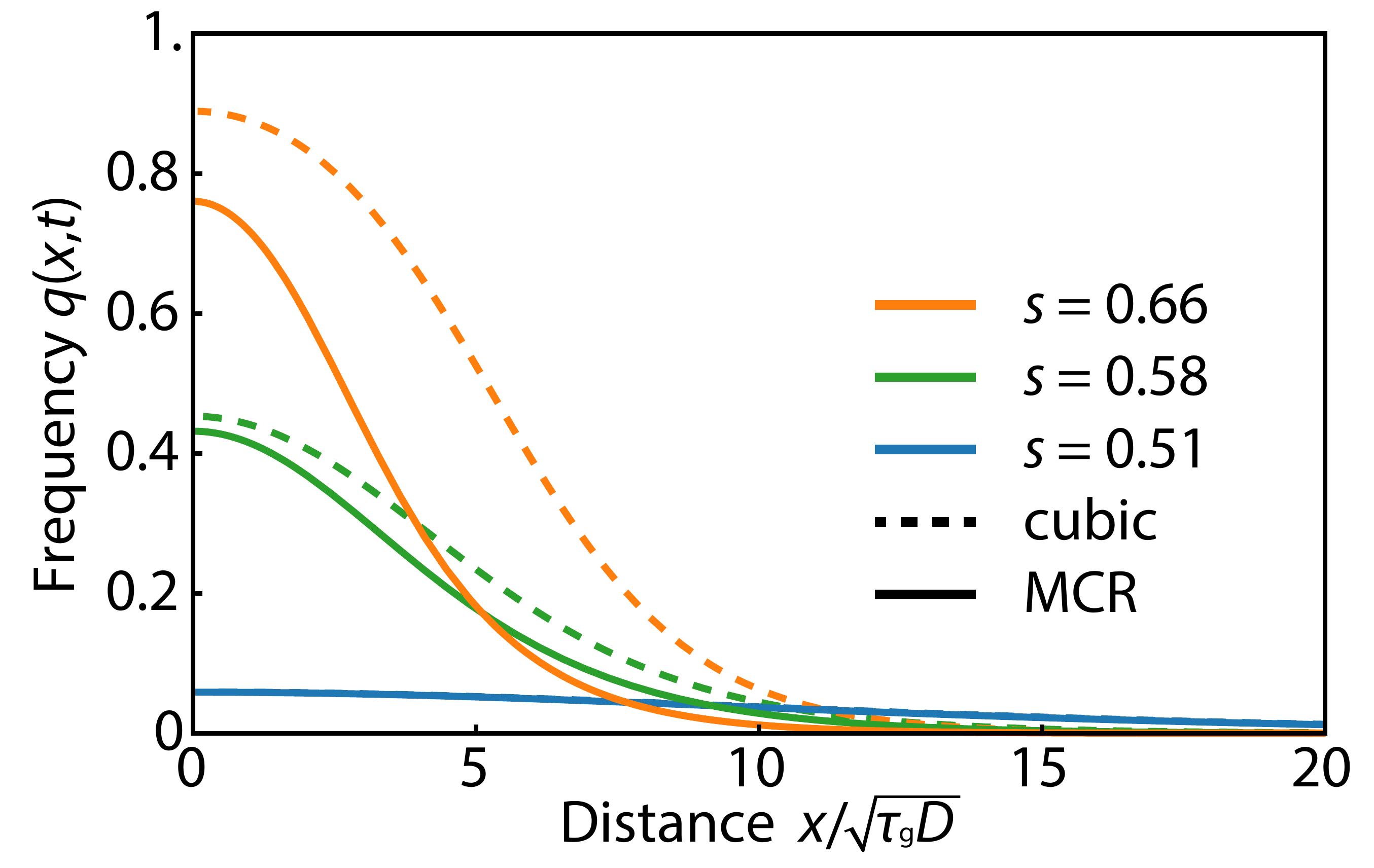}
\caption{
Initial critical frequency profiles of the mutagenic chain reaction (MCR) allele $q_{\rm{c}}(x)$ just sufficient to excite a pushed genetic wave in 1D (critical propagule). 
Numerically calculated critical propagules for the MCR model of Eq.~\ref{rdMCR} (solid lines) are compared with analytical results available for the cubic model Eq.~\ref{Fcubic} (dashed lines) \cite{barton2011spatial}.
When $s=0.51$, the two equations gives almost identical results, but as $s$ increases the critical propagule shape of the MCR model deviates significantly from that of the cubic model. The critical propagule of the cubic equation consistently overestimates the height of the $q_{\rm{c}}(x)$, since the $sq(2-q)>0$ term in the denominator of the MCR model always increases the growth rate. 
}
\label{Fig_5}
\end{figure} 

We show the spatial concentration profile $q_{\rm{c}}(x)$ that constitutes that (Gaussian) critical nucleus just sufficient to initiate an excitable wave in Fig.~\ref{Fig_5}.
The solid lines represent numerically obtained critical nuclei of the MCR model.
Note the consistency for $s=0.58$ with the pushed excitable waves shown in Fig.~\ref{Fig_4}.
The dashed lines represent analytically derived critical propagules of the cubic model as a reference  (see SI Appendix for details).
Fig.~\ref{Fig_5} shows that the cubic model overestimates the height of critical propagule, particularly for larger $s$.
The difference between the reaction terms of the MCR model $f_{\rm{MCR}}(q)$ (see Eq.~\ref{FMCR}) and that of its cubic approximation $f_{\rm{cubic}}(q)$ (see Eq.~\ref{Fcubic}), arises from the term $-sq(2-q)$ in the denominator of Eq.~\ref{rdMCR}. 
In the biologically relevant regime $(0<s<1,~0<q<1)$, $sq(2-q)$ is always positive and $f_{\rm{MCR}}(q)>f_{\rm{cubic}}(q)$ is satisfied, which explains why there is a larger critical propagule in the cubic approximation, and the discrepancy is larger for larger $s$.
The critical nucleus with a step-function-like circular boundary is studied both numerically and analytically in two dimensions in the SI Appendix.

\subsection*{Stopping of pushed, excitable waves by a selective disadvantage barrier} 
Thus far, we have found that
(i)~we can control initiation of the spatial spread of a gene drive provided $s_{\rm{min}}=0.5<s<s_{\rm{max}}=0.697$, 
and (ii)~the pushed population waves in this regime slow down and eventually stop (and reverse direction) when $s>s_{\rm{max}}$, see SI Appendix.
In this section, we examine alternative ways to confine an excitable gene drive wave to attain greater control over its spread in this regime.

Imagine exploiting the CRISPR/Cas9 system to encode multiple functionalities into the gene drive machinery \cite{cong2013multiplex, jinek2013rna,mali2013rna,gantz2015mutagenic}. 
For example, one could produce genetically engineered mosquitoes that are not only resistant to malaria, but also specifically vulnerable to an insecticide that is harmless for the wild-type alleles.
Such a gene drive, which is uniquely vulnerable to an otherwise harmless compound, is a sensitizing drive \cite{esvelt2014concerning}.
The effect of laying down insecticide in a prescribed spatial pattern on a sensitizing drive can be incorporated in our model by increasing the selective disadvantage to a value $s_b(>s)$ within a ``selective disadvantage barrier'' region.

\begin{figure}[b!]
\centering
\includegraphics[clip,width=0.8\columnwidth]{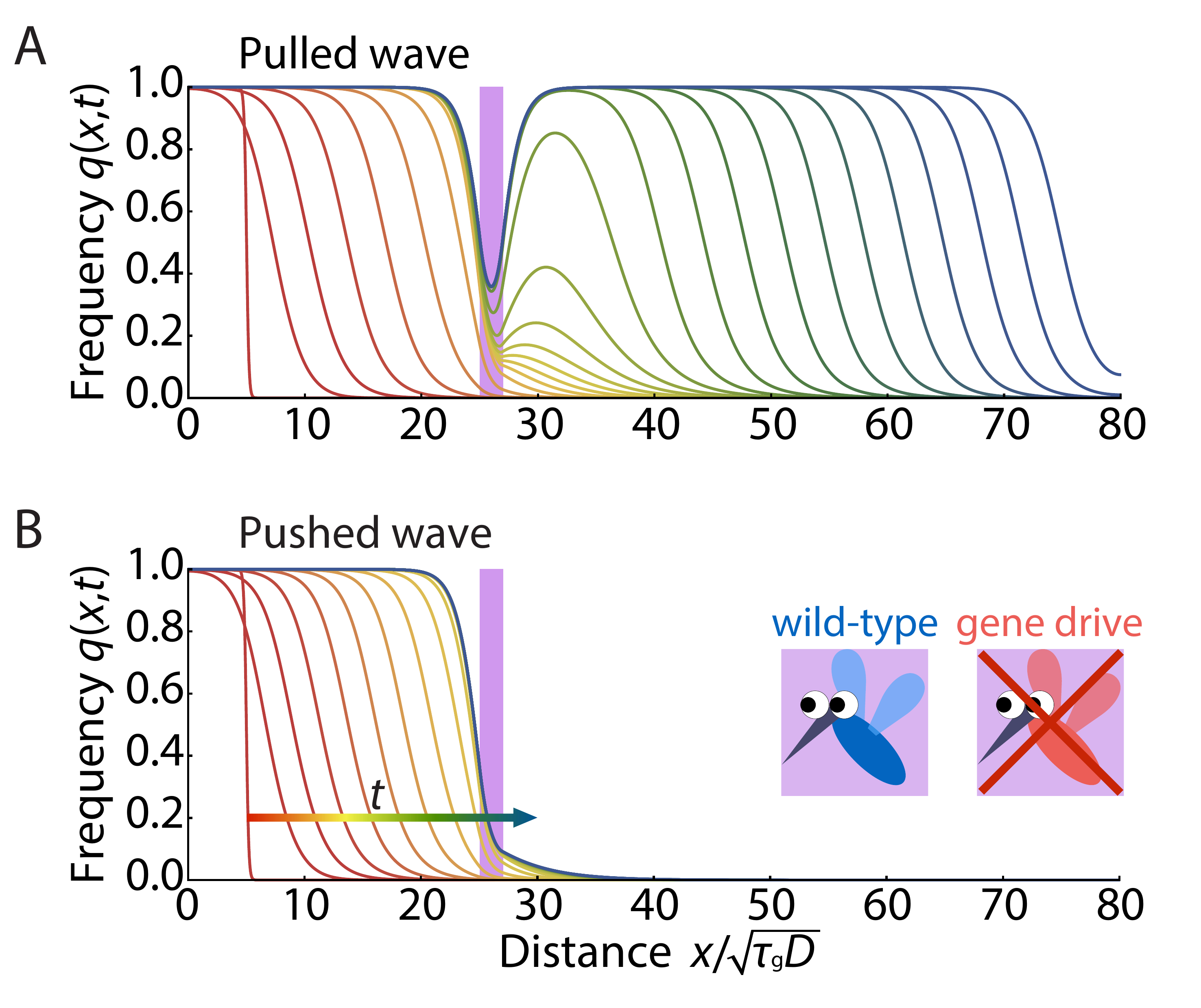}

\caption{
Numerical simulations of pushed, excitable waves generated by Eq.~\ref{rdMCR}
 with barriers in one dimension, with time increments $\Delta t = 5.0 \tau_g$. 
As the waves advance from left to right, the early time response is shown in red with later times in blue.
The fitness disadvantage inside the barrier is set to $s_b=0.958$ within a region $25\sqrt{\tau_g D}<x<27\sqrt{\tau_g D}$ (shown as a purple bar).
The initial conditions are step-function-like, $q(x,0)=q_0/(1+e^{10(x-x_0)/\sqrt{\tau_g D}})$, with $q_0 = 1.0$ and $x_0=5.0\sqrt{\tau_gD}$, similar to the initial condition Eq.~S30 we used in two dimensions (see SI Appendix).
(A) In the case of a Fisher wave with $s =0.479 <s_{\rm{min}}=0.5$, a small number of individuals diffuse through the barrier, which is sufficient to reestablish a robust traveling wave.
(B) In the case of the excitable wave $s =0.542>s_{\rm{min}}=0.5$, a small number of individuals also diffuse through the barrier. However, since the tail of the penetrating wave front is insufficient to create a critical nucleus, the barrier causes the excitable wave to die out.
}
\label{Fig_6}
\end{figure} 
In Fig.~\ref{Fig_6}, we numerically simulate the mutagenic chain reaction model defined by Eq.~\ref{rdMCR} in one dimension with a barrier of strength $s_b=0.958$ placed in a region $25\sqrt{\tau_g D}<x<27\sqrt{\tau_g D}$.
When the selective disadvantage outside the barrier is small $(s<0.5)$ and the population wave travels as the pulled Fisher wave, even a tiny fraction of MCR allele diffusing through the insecticide region can easily reestablish the population wave, as shown in Fig.~\ref{Fig_6}\emph{A}.
However, when the system is in the pushed wave regime $0.5<s<0.697$, the wave can be stopped provided the spatial profile of the gene drive allele that leaks through does not constitute a critical nucleus, as illustrated in Fig.~\ref{Fig_6}\emph{B}.
See the SI Appendix for numerically calculated plots of the critical width and barrier selective disadvantage needed to stop pushed waves for various values of $s$.

\subsection*{Excitable Wave Dynamics with Gapped Barriers in Two Dimensions}
\begin{figure}[htp]
\centering
\includegraphics[clip,width=0.9\columnwidth]{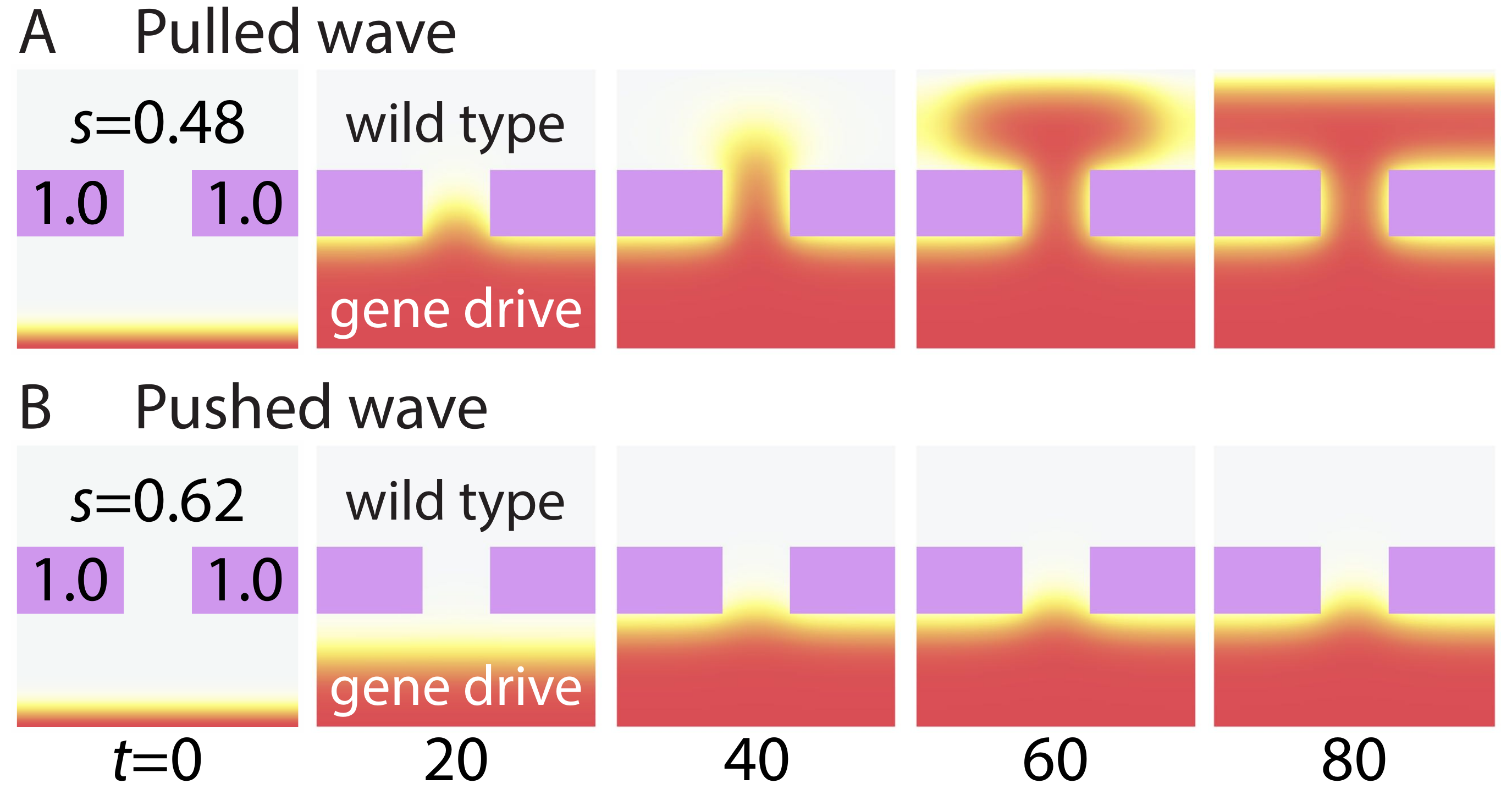}
\caption{
Population waves impeded by a selective disadvantage barrier of strength $s_b = 1.0$ (colored purple) with a gap.
This imperfect barrier has a region without insecticide in the middle of width $6\sqrt{\tau_g D}$.
 (A) The pulled Fisher wave with $s=0.48<0.5$ always leaks through the gap and reestablishes the gene drive wave (colored red and yellow). (B) The pushed wave that arises when $s=0.62>0.5$ is deexcited by a gapped barrier, provided the gap width is comparable to or smaller than the width of the gene drive wave.
}
\label{Fig_7}
\end{figure}

In the previous section, we showed that pushed excitable waves can be stopped by a selective disadvantage barrier in one dimension. However, in two dimensions, it may be difficult to make barriers without defects.
Hence, we have also studied the effect of a gap in a two-dimensional selective disadvantage barrier.
We find that while the gene drive population wave in the Fisher wave regime $s<0.5$ always leaks through the gaps,
the excitable wave with $0.5<s<0.697$ can be stopped, provided the gap is comparable or smaller than the width of the traveling wave front.
In Fig.~\ref{Fig_7}, we illustrate the gene drive dynamics for two different parameter choices.
Both in Fig.~\ref{Fig_7}\emph{A} and \emph{B}, the strength of the selective disadvantage barrier is set to be $s_b=1.0$ and the width of the gap in the barrier is set to be $6\sqrt{\tau_g D}$.
The engineered selective disadvantage in the non-barrier region $s$ differs in the two plots.
In Fig.~\ref{Fig_7}\emph{A} $s=0.48<0.5$, so the gene drive wave propagates as a pulled Fisher wave and the wave easily leaks through the gap.
If genetic drift can be neglected, we expect that Fisher wave excitations will leak through any gap however small.
However, when the selective disadvantage barrier is in the pushed wave regime $0.5<s<0.697$, the population wave can be stopped by a gapped selective disadvantage barrier as shown in Fig.~\ref{Fig_7}\emph{B}. 
To stop a pushed excitable wave, the gap dimensions must be smaller than the front width; alternatively, we can say that the gap must be smaller than size of the critical nucleus.

\subsection*{Discussion}
The CRISPR/Cas9 system has greatly expanded the design space for genome editing and construction of mutagenic chain reactions with non-Mendelian inheritance.
We analyzed the spatial spreading of gene drive constructs, applying reaction-diffusion formulations that have been developed to understand spatial genetic waves with bistable dynamics \cite{barton1979dynamics, barton1989adaptation, barton2011spatial}. For a continuous time and space version of the model of Unckless \emph{et al} \cite{unckless2015modeling},
in the limit of $100\%$ conversion efficiency, we found that a critical nucleus or propagule is required to establish a gene drive population wave when the selective disadvantage satisfies $0.5<s<0.697$. 
Our model led us to study termination of pushed gene drive waves using a barrier that acts only on gene drive homozygotes, corresponding to an insecticide in the case of mosquitoes. 
In this parameter regime, the properties of pushed waves allow safeguards against the accidental release and spreading of the gene drives.
One can, in effect, construct switches that initiate and terminate the gene drive wave.
In the future, it would be interesting to study the stochasticity due to finite population size (genetic drift), which is known to play a role in the first quadrant of Fig.~\ref{Fig_2} \cite{korolev2011competition, lavrentovich2014asymmetric}.
We expect that genetic drift can be neglected provided $N_{\rm{eff}} \gg 1$, where $N_{\rm{eff}}$ is
an effective population size, say, the number of organisms in a well-mixed critical propagule. See the SI Appendix for a brief discussions on genetic drift.
It could also be important to study the effect of additional mutations on an excitable gene drive wave, particularly those that move the organism outside the preferred range $0.5<s<0.697$.
Finally we address possible experimental tests of the theoretical predictions.
Since it seems inadvisable to conduct field tests without thorough understanding of the system, laboratory experiments with microbes would be a good starting point.
Recently, the transition from pulled to pushed waves was qualitatively investigated with haploid microbial populations \cite{gandhi2016range}. Because the mutagenic chain reaction has already been realized in \emph{S. Cerevisiae} \cite{dicarlo2015safeguarding},
it may also be possible to test the theory in the context of range expansions on a Petri dish, as has already been done for haploid mutualistic yeast strains in \cite{muller2014genetic}.
Here, the frontier approximates a one dimensional stepping stone model, and jostling of daughter cells at the frontier leads to an effective diffusion constant in one dimension \cite{RevModPhys.82.1691, hallatschek2007genetic}.
Finally, as illustrated in Fig.~S2, the mathematics of the spatial evolutionary games in one dimension parallels the dynamics of diploid gene drives in the pushed wave regime, providing another arena for experimental tests, including the effects of genetic drift.

\subsection*{Numerical Simulations}
To simulate the dynamics governed by Eq.~\ref{FMCR} in Figs.~\ref{Fig_4},\ref{Fig_6},\ref{Fig_7} and S6, we used the method of lines and discretized spatial variables to map the partial differential equation to a system of coupled ordinary equations (``ODE''). Then we solved the coupled ODEs with a standard ODE solver. The width of the spatial grids were varied from $\frac{1}{200}\sqrt{\tau_g D}$ to $\frac{1}{20}\sqrt{\tau_g D}$ always making sure that the mesh size was much smaller than the width of the fronts of the pushed and pulled genetic waves we studied.

\subsection*{Acknowledgement}
We thank N. Barton, S. Block, S. Sawyer, T. Stearns, and M. Turelli for helpful discussions and two anonymous reviewers for useful suggestions. N. Barton also provided a critical reading of our manuscript. 
Work by HT and DRN was supported by the National Science Foundation, through grants DMR1608501 and via the Harvard Materials Science Research and Engineering Center via grant DMR1435999. HAS acknowledges support from NSF grants MCB1344191 and DMS1614907.

\clearpage

\renewcommand{\theequation}{S\arabic{equation}}    
\setcounter{equation}{0}   
\setcounter{figure}{0}  
\setcounter{page}{1}
\captionsetup[figure]{name=Figure}
\captionsetup[table]{name=Table}
\renewcommand\thefigure{S\arabic{figure}}
\renewcommand\thetable{S\arabic{table}}

\onecolumngrid
\section*{Supporting Information (SI)}
\titlespacing\section{3pt}{10pt plus 4pt minus 2pt}{0pt plus 2pt minus 2pt}
\titlespacing\subsection{3pt}{10pt plus 4pt minus 2pt}{0pt plus 2pt minus 2pt}
\titlespacing\subsubsection{3pt}{10pt plus 4pt minus 2pt}{0pt plus 2pt minus 2pt}

\tableofcontents
\FloatBarrier\subsection{Nucleation theory of the gene drive population waves}
Here we identify different parameter regimes of various types of gene drive waves by establishing an analogy between zero temperature nucleation theory and the reaction-diffusion equation of the prescribed mutagenic chain reaction,
\begin{equation} \label{rdMCR2}
\frac{\partial q}{\partial t} = D \frac{\partial^2 q}{\partial x^2}+ \frac{1}{\tau_g} \frac{sq(1-q)(q-q^*)}{1-sq(2-q)},
\end{equation} 
using the methods reviewed in \cite{barton2011spatial}.
First, we introduce a potential energy function $U(q)$
\begin{equation}
U(q) = -\frac{1}{\tau_g} \int^{q}_{0} \frac{sq'(1-q')(q'-q^*)}{1-sq'(2-q')} dq', q^* = \frac{2s-1}{s},
\end{equation}
and rewrite Eq.~\ref{rdMCR2} as
\begin{equation}
\frac{\partial q}{\partial t} = D \frac{\partial^2 q}{\partial x^2} - \frac{dU(q)}{dq}.
\end{equation}
It is useful to recast the reaction-diffusion dynamics in terms of a functional derivative
\begin{equation}
\frac{\partial q(x,t)}{\partial t} = -\frac{\delta \mathcal{F}[q(y,t)] }{\delta q(x,t)},
\end{equation}
where the functional $\mathcal{F}[q(y,t)]$ is given by
\begin{equation}
\mathcal{F}[q(y,t)] = \int^{\infty}_{-\infty} \bigg \{ \frac{1}{2} D \Big( \frac{\partial q(y,t)}{ \partial y} \Big)^2 +U[q(y,t)] \bigg \} dy,
\end{equation}

and we have
\begin{equation}
\begin{split}
&-\frac{\delta \mathcal{F}[q(y,t)] }{\delta q(x,t)}
= - \lim_{\epsilon \rightarrow 0} \frac{ \mathcal{F}[q(y,t) + \epsilon \delta(y-x)] - \mathcal{F}[q(y,t)] }{\epsilon}\\
&= -   \int_{-\infty}^{\infty} 
\Big\{ D \frac{\partial q(y,t)}{\partial y} \frac{\partial \delta (y-x)}{\partial y} + \frac{dU[q(y,t)]}{dq} \delta(y-x) \Big\} dy\\
&= D \frac{\partial^2 q(x,t)}{\partial x^2}  - \frac{dU[q(x,t)]}{dq}.
\end{split}
\end{equation}

Since $\mathcal{F}(t)$ always decreases in time, 
\begin{equation}
\begin{split}
\frac{d \mathcal{F}(t)}{dt}
&=\int_{-\infty}^{\infty} \frac{\partial q(x,t)}{\partial t} \frac{\delta \mathcal{F}[q(y,t)]}{\delta q(x,t)}dx\\
&= -  \int_{-\infty}^{\infty}\Big( \frac{\partial q(x,t)}{\partial t} \Big)^2 dx \leq 0,
\end{split}
\end{equation}
$\mathcal{F}[q(y,t)]$ plays the role of the free energy in a thermodynamic system.

The potential energy function $U(q)$ with various selective disadvantages $s$ is plotted in Fig.~3. 
$U(1)$ becomes the absolute minimum when $0.5<s$ and population waves behave as pushed waves, because both $U(0)$ and $U(1)$ are locally stable \cite{barton1979dynamics, barton1989adaptation, barton2011spatial}. The pushed gene drive wave stalls out when the two stable points have the same potential energy (blue curve in Fig.~3).
The maximum value of the selective disadvantage $s_{\rm{max}}$ supporting the pushed wave of the gene drive allele can be derived by equating $U(0)=U(1)$, which leads to
\begin{equation}
\begin{split}
0 &=\int^{1}_{0} \frac{sq(1-q)(q-q^*)}{1-sq(2-q)}dq\\
&=\frac{-2+s_{\rm{max}}+2\sqrt{-1+\frac{1}{s_{\rm{max}}}}\arcsin(\sqrt{s_{\rm{max}}})}{2s_{\rm{max}}}.\\
&\Rightarrow s_{\rm{max}} \approx 0.697
\end{split}
\end{equation}
The excitable gene drive wave of primary interest to us thus arises when the selective disadvantage satisfies
\begin{equation}
0.5 < s < 0.697.
\end{equation}

\subsection{The range of the pushed wave regime with an arbitrary conversion rate}
\begin{figure}[htp]
\centering
\includegraphics[clip,width=0.8\columnwidth]{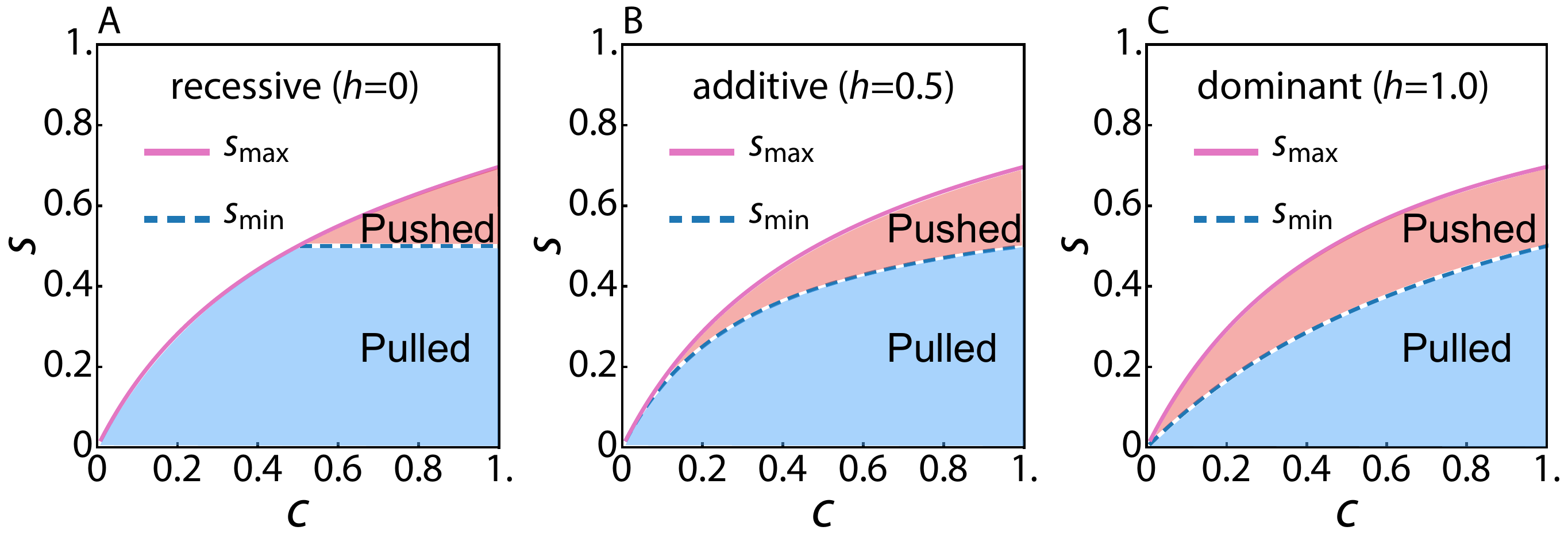}
\caption{
$s_{\rm{min}}$ and $s_{\rm{max}}$ as a function of the conversion rate $c$ when the fitness of heterozygotes individuals is (A) recessive ($h=0$), (B) additive ($h=0.5$) and (C) dominant ($h=1.0$) of gene drives, where the fitness of heterozygotes is $1-hs$.
The socially responsible pushed wave regime ($s_{\rm{min}}<s<s_{\rm{max}}$) is always widest when $c=1$, i.e., for $100\%$ conversion efficiency. Note that the results become independent of $h$ when $c=1$. The gene drive wave reverses direction and dies out in the white regions of this diagram.
}
\label{Fig_S1}
\end{figure} 
In the main text, we assumed perfect conversion efficiency ($c=1$) of the mutagenic chain reaction. However, in reality, some fraction of the reactions can be unsuccessful and the conversion rate $c$ will be $0<c<1$.
As a result there will be heterozygous individuals with fitness $1-hs$, where $h$ controls dominance of the gene drive allele. 
When $h=1$, the gene drive allele is dominant and the fitness of the heterozygous genotype is $1-s$. The choices $h=0, 0.5$ correspond to the recessive and additive cases respectively.
As derived by Unckless \emph{et al.} \cite{unckless2015modeling}, the reaction term in Eq.~5 is now given by

\begin{equation}
\begin{split}
&q(t+\tau_g) - q(t)
=\bar{f}(q)\\
&=\frac{q^2(1-s)+q(1-q)\big[ (1-c) (1-hs) + 2c(1-s) \big]}{q^2 (1-s) +2q(1-q)(1-c)(1-hs)+2q(1-q)c(1-s)+(1-q)^2} - q
\end{split}
\end{equation}
There are again three fixed points $q=0,1,q^*$ where the third fixed point is
\begin{equation}
q^*=\frac{c+cs(h-2)-hs}{s(1-2c-2h+2ch)}.
\end{equation}
Following \cite{unckless2015modeling}, we find that $q^*$ first becomes positive for $s>s_{\rm{min}}$, where
\begin{equation}
s_{\rm{min}} = \frac{c}{2c - (c-1)h}.
\end{equation}
For $0\leq s \leq s_{\rm{min}}$, $q^*<0$ and the spatial dynamics is again controlled by pulled waves.
We can also calculate $s_{\rm{max}}$ by recalculating the potential function analogy discussed in SI, Sec~\textbf{A} and in the main text,
\begin{equation}
\bar{U}(q)= -\frac{1}{\tau_g} \int_0^q \bar{f}(q') dq',
\end{equation}
and numerically solving for $\bar{U}(q=0,c,h,s_{\rm{max}})=\bar{U}(q=1,c,h,s_{\rm{max}})$ to obtain $s_{\rm{max}}(c)$ given $h$,
 with the results shown in Fig. S1. The gene drive spreads spatially as a pushed excitable wave for $s_{\rm{min}} < s < s_{\rm{max}}$. Note that the relevant range of $s$ when $c < 1$ shrinks compared to $c=1$.

\FloatBarrier\subsection{Spatial evolutionary games in one dimension}
\begin{figure}[htp]
\centering
\includegraphics[clip,width=0.5\columnwidth]{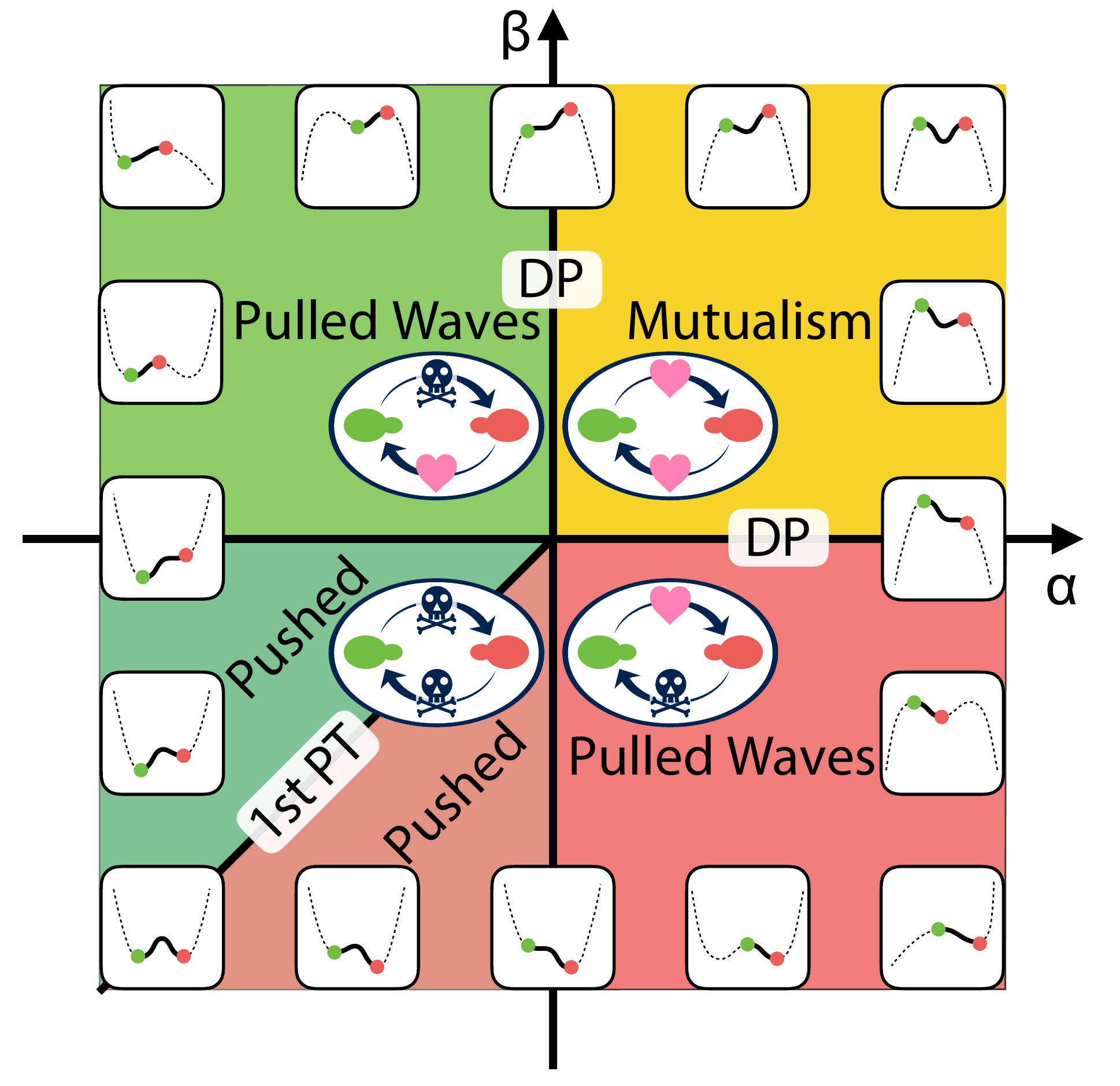}
\caption{ 
A schematic phase diagram of the spatial evolutionary games in one dimension ignoring genetic drift.  
The parameters $\alpha$ and $\beta$ describe interactions between red and green genetic variants, with growth rates written as $w_R (x,t) = g + \alpha (1-f(x,t))$ and $w_G (x,t) = g + \beta f(x,t)$ respectively.
(The parameter $g>0$ is a background growth rate.)
Inserted graphs show schematically the potential energy function $U(f)$, where each of the green and red dot corresponds to $f=0$ and $f=1$ respectively ($0 \leq f \leq 1$). 
By searching for barriers in $U(f)$ as a function of $\alpha$ and $\beta$, we identify the bistable regimes that require a critical nucleus and pushed excitable waves to reach a stable dynamical state and the pulled Fisher wave regimes which do not require the nucleation process.
The two regimes are separated by two solid black lines $\alpha=0$, $\beta<0$, and $\alpha<0$, $\beta=0$, which correspond limits of metastability. 
The solid line along $\alpha=\beta<0$ between the two bistable states is analogous to a first-order phase transition line (equal depth minima in $U(q)$),
along which the excitable genetic wave separating red and green stalls out. 
}
\label{Fig_S2}
\end{figure} 
In this SI section, we show that genetic waves mathematically quite similar to the pushed gene drive waves studied here arise in spatial evolutionary games of two interacting asexual species that are colored red (``$R$'') and green (``$G$'') using the analogy with nucleation theory introduced in the previous SI section. 
We start from the continuum description of the one dimensional stepping stone model (following \cite{RevModPhys.82.1691,korolev2011competition}), 
\begin{equation}
\frac{\partial f(x,t)}{\partial t} = D \frac{\partial^2 f(x,t)}{\partial x^2} + s[f]f(1-f) + \sqrt{D_g f(1-f)} \Gamma (x,t),
\label{FullRG}
\end{equation}
where $f(x,t)$ is the frequency of red species and $D$ is the spatial diffusion constant representing migration.
The last term, where $\Gamma(x,t)$ is an Ito correlated Gaussian white noise source and $D_g$, proportional to an inverse effective population size, represents genetic drift. We henceforth neglect genetic drift and set this term to zero.
The function $s[f]$ represents the difference in relative reproduction rates between the two species, and is given by \cite{RevModPhys.82.1691}
\begin{equation}
s[f]=w_{\rm{eff}} = \frac{w_R - w_G}{\frac{1}{2} (w_R + w_G)},
\end{equation}
where $w_R$ and $w_G$ are fitnesses of alleles $R$ and $G$. If $g$ is a background reproduction rate, we have
\begin{equation}
\begin{split}
w_R (x,t) &= g + \alpha (1-f(x,t)),\\
w_G (x,t) &= g + \beta f(x,t),
\end{split}
\end{equation}
where the interactions between the two competing variants are characterized by constants $\alpha$ and $\beta$.
With the definitions above, we have 
\begin{equation}
s[f]=-\frac{(\alpha + \beta)(f-\frac{\alpha}{\alpha+\beta})}{g+\frac{1}{2}\alpha (1-f) + \frac{1}{2} \beta f},
\end{equation}
which leads to a reaction term similar to that in Eq.~5 and introduces an additional fixed point into the dynamics of Eq.~\ref{FullRG} at 
$f^* = \frac{\alpha}{\alpha + \beta}$ in addition to $f=0,1$.
A diagram summarizing the dynamics of this model is shown in Fig.~2.
This ``phase diagram'' was worked out including genetic drift in Eq.~\ref{FullRG} which affects the shape and location of the phase transition lines in the first quadrant of Fig.~1. \cite{korolev2011competition}.
If the genetic drift term in Eq.~\ref{FullRG} is neglected, the lines labelled ``DP'' in Fig. 2 would coincide with the positive $\alpha$ and $\beta$ axes and would merge at the origin.
Upon setting $D_g = 0$ in Eq.~\ref{FullRG}, we employ the argument presented above and define a potential energy function,   
\begin{equation}
U_b (f)= -\int^{f}_0 s[f']f'(1-f') df'.
\label{Ub}
\end{equation}

The schematic picture of $U_b (f)$ in different parameter regimes is drawn in Fig.~\ref{Fig_S2}.
The mutualistic regime ($\alpha>0, \beta>0$) has already been studied in detail, 
including effects of genetic drift \cite{korolev2011competition}.
By studying shapes of the potential energy function $U[f]$ we identify two important parameter regimes.
In the bistable regime (dark green), there is a finite energy barrier between the two locally stable states and a nucleation process is required to establish an excitable wave.

However, in the Fisher wave regimes (light green and light red), there is no energy barrier to reach the unique stable configuration and thus nucleation is not required.
The two regimes are separated by the two black solid lines $\alpha=0, \beta<0$ or $\alpha<0, \beta=0$,
which are limits of metastability.
We also draw a solid black line between the two bistable states along $\alpha=\beta<0$,
where the pushed waves stall out. This line is analogous to a line of first-order transitions.
When $\alpha \neq \beta$, the integral in Eq.~\ref{Ub} for the effective thermodynamic potential is given by

\begin{equation}
\begin{split}
U[f] &=
 \frac{1}{3(\alpha-\beta)^4}\Bigg( (\alpha-\beta) f \bigg\{ \alpha^3 f (2f-3)
  + \alpha^2 f (9\beta - 2\beta f +6g)\\
&+ \alpha \Big(\beta^2 \big(12-f(3+2f) \big) + 36\beta g+24g^2 \Big)
+\beta \big( \beta^2 f(-3+2f) - 6\beta(-2+f)g+24g^2 \big) \bigg\}\\
&+ 12(\alpha+2g)(\beta+2g) \big( \alpha \beta + (\alpha + \beta)g \big) \log \Big[1-\frac{\alpha-\beta}{\alpha+2g} f \Big]
\Bigg).
\label{long}
\end{split}
\end{equation}

When $\alpha=\beta$, we can simplify $s[f]$ 
\begin{equation}
s[f] = -\frac{2 \alpha (f-\frac{1}{2})}{g+\frac{1}{2}\alpha},
\end{equation}
and the integral gives
\begin{equation}
U[f] = \frac{2 \alpha }{g+\frac{1}{2}\alpha} \int^{f}_0 f'(1-f') \Big( f'-\frac{1}{2} \Big) df' =-\frac{\alpha}{2g+\alpha} {f}^2 ({f}-1)^2.
\end{equation}

When $\alpha=-\beta$, $\alpha \ll g$ and $1 \ll \big| \frac{g}{\alpha}+\frac{1}{2} \big|$, we have 
\begin{equation}
s[f] = \frac{\alpha}{g+\frac{1}{2}\alpha (1-2f)}
\end{equation}
and
\begin{equation}
\begin{split}
&U[f] = -\int^{f}_0 \frac{f'^2 -f'}{f' - \big( \frac{g}{\alpha}+\frac{1}{2} \big)} df'\\
&= -\frac{1}{2}f \Big(f+ \frac{2g}{\alpha} -1 \Big) - \bigg(\frac{g}{\alpha} + \frac{1}{2} \bigg) \bigg(\frac{g}{\alpha} - \frac{1}{2} \bigg) \log \bigg[ 
1-\frac{2\alpha f}{\alpha + 2g}
\bigg]
\label{amib}
\end{split}
\end{equation}
The last term diverges at $f=\frac{g}{\alpha}+\frac{1}{2}$, but we focus on the weak interaction limit $1 \ll \big| \frac{g}{\alpha}+\frac{1}{2} \big|$, where the biologically relevant regime $0 \leq f \leq 1$ will not be affected.
If we substitute $\alpha=-\beta$ into Eq.~\ref{long}, we recover Eq.~\ref{amib}, as expected.

\FloatBarrier
\subsection{Calculation of the critical propagules in one dimension}
In this SI section, we describe details of the calculation of the critical propagules shown in Fig.~5. 
Reaction-diffusion equations in one dimension with a general reaction term $R[q(x,t)]$ can be written as
\begin{equation}\label{general}
\tau_g \frac{\partial q(x, t)}{\partial t} = \tau_g D \frac{\partial^2 q(x,t)}{\partial x^2} + R[q(x,t)].
\end{equation}
The critical propagule profile $q_{\rm{c}}(x)$ can be defined as a stationary solution of Eq.~\ref{general}, i.e.,
\begin{equation}
0 = \tau_g D \frac{\partial^2 q_c}{\partial x^2} + R[q_c].
 \label{ODE}
\end{equation}
Upon multiplying both sides by $\frac{d q_{\rm{c}}}{d x}$ and integrating we obtain,
\begin{equation}
\tau_g D \Big ( \frac{d q_{\rm{c}}}{d x} \Big)^2 =2 \int^0_{q} R[\tilde{q}] d\tilde{q}.
\end{equation}
If we assume a symmetric critical propagule about $x=0$, so that $\frac{d q_{\rm{c}}}{d x} = 0$ at $x=0$, we can obtain $q_{\rm{m}} \equiv q_{\rm{c}}(0) $ from

\begin{equation}
\int^0_{q_{\rm{m}}} R[\tilde{q}] d\tilde{q}=0.
\end{equation}
Since the slope $\frac{dx_{\rm{c}} (q)}{dq}$ is given by
\begin{equation}
\frac{d x_{\rm{c}}(q)}{d q} = \frac{\sqrt{\frac{\tau_g D}{2}}}{\sqrt{ \int^0_{q} R[\tilde{q}] d\tilde{q}}},
\end{equation}
we obtain the critical propagule profile $x_{\rm{c}}(q)$ by integrating both sides from $q_{\rm{m}}$ to $q$.
The calculations described above can be carried out analytically for the cubic reaction term Eq.~7 and critical propagules for $s=0.66,0.58,0.51$ are plotted in Fig.~5 with dashed lines.
For the full MCR equation, the corresponding numerical results are plotted with solid lines.

\FloatBarrier
\subsection{Critical radius and allele concentration in two dimensions}
\begin{figure} 
\centering
\includegraphics[clip,width=0.5\columnwidth]{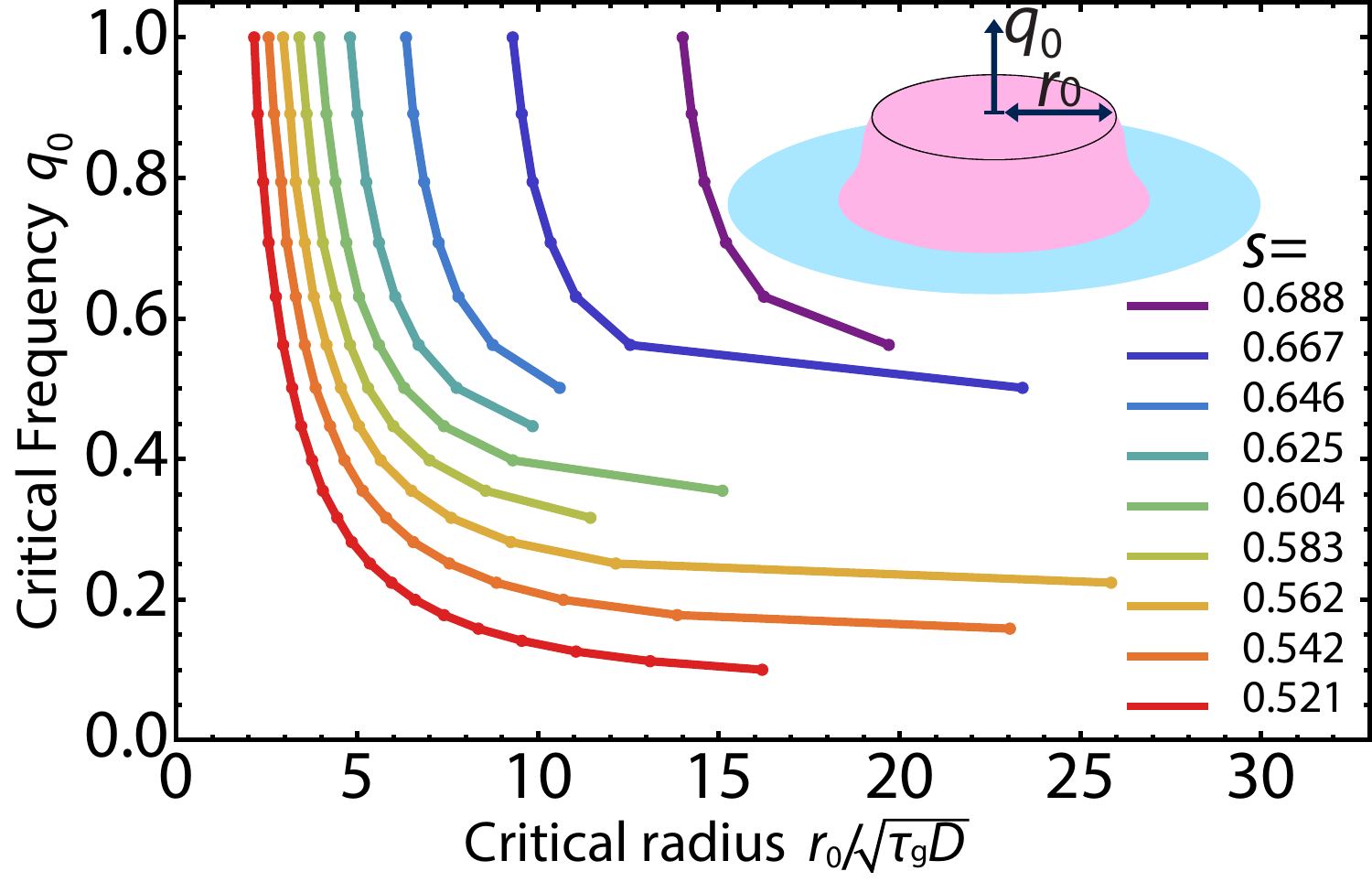}
\caption{
In two dimensions the gene drive allele is introduced uniformly over a disk-shaped region with radius $r_0$ with uniform frequency $q_0$ inside as illustrated in the inset image.
We numerically determined the critical frequency $q_0$ and radius $r_0$ just sufficient to initiate an excitable wave in two dimensions.}
\label{Fig_S3}
\end{figure} 

In practice, it is important to model the distribution of MCR alleles to be released locally to initiate its traveling genetic wave in a two-dimensional space.
Upon assuming circular symmetry of the traveling wave solution, the reaction-diffusion equation governing the radial frequency profile of the MCR allele $q(r,t)$ reads in radial coordinates,
\begin{equation}
\tau_g \frac{\partial q}{\partial t}=  \tau_g D \Big( \frac{\partial^2 q}{\partial r^2}+\frac{1}{r} \frac{\partial q}{\partial r} \Big) + \frac{sq(1-q)(q-q^*)}{1-sq(2-q)}.
\end{equation}
The only correction to the one dimensional case is the derivative term $\frac{1}{r} \frac{\partial q}{\partial r}$, which can be neglected relative to $\frac{\partial^2 q}{\partial r^2}$ in the limit of $r\rightarrow \infty$.
However, we keep this term in the calculation of the critical nucleus as this term is not negligible where $r$ is comparable to or smaller than the width of the excitable wave being launched.
In our numerical calculations, instead of a Gaussian initial condition,
it is convenient to introduce the gene drive allele with a uniform frequency $q_0$ over a circular region with radius $r_0$.
Indeed, in an actual release of a gene drive organism, it is plausible that the release would be implemented by creating a gene drive concentration $q_0$ in a circular region of radius $r_0$ with a sharp boundary.
To model the radial frequency profiles, we used a circularly symmetric steep logistic function as an initial condition,
\begin{equation}\label{logistic}
q(r,t=0)=\frac{q_0}{1+e^{10(r-r_0)/\sqrt{\tau_g D}}},
\end{equation}
instead of a step function to insure numerical stability.
Fig.~\ref{Fig_S3} shows the parameter regimes where a pushed wave is excited for various selective disadvantages $s$. The pushed waves successfully launched for initial conditions whose parameters are above the curves $q_0(r_0)$, shown for a variety of selective disadvantages $s$ in the pushed wave regime.

\FloatBarrier
\subsection{Line tension, energy difference and analogy with nucleation theory in two dimensions}
The scenario studied in the previous section (sharp boundary, adjustable initial drive concentration $q_0$ and inoculation radius $r_0$) seems appropriate for many engineered releases of gene drives, at least in situations with large effective population sizes $N_{\rm{eff}}$, so that genetic drift can be neglected. (See the discussion of genetic drift in SI Sec.~\textbf{J}.)

However, when genetic drift is important, stochastic contributions like the term $\sqrt{D_g f(1-f) \eta(x,t)}$ in, e.g., Eq. \ref{FullRG}, can act on spatial gradients at the interfaces of pushed and pulled waves \cite{polechova2011genetic, polechova2015limits} in a manner somewhat reminiscent of thermal fluctuations near a first-order phase transition. Provided strong genetic drift is able to produce something analogous to local thermal equilibrium after a gene drive release, it is interesting to explore an analogy with classical nucleation theory.
Nucleation leads to a pushed wave when $s_{\rm{min}}<s<s_{\rm{max}}$. One might then expect the two-dimensional analog of the total energy function discussed in SI Sec.~\textbf{A} for an equilibrated circular droplet with $q_0=1$ and radius $r_0$ in two dimensions to take the form  
\begin{equation}
\begin{split}
\mathcal{F}[q(\bm{r})] &= \int d \bm{r} \bigg\{ \frac{1}{2} D \big( \bm{\nabla} q(\bm{r}) \big)^2 + U[q(\bm{r})] \bigg\}\\
&= 2\pi \int_0^{\infty} dr \frac{rD}{2} \Big( \frac{dq}{dr} \Big)^2 + 2 \pi \int_0^{\infty} dr r U[q(r)]\\
&\approx 2\pi r_0  \int_0^{\infty} dr \frac{D}{2} \Big( \frac{dq}{dr} \Big)^2 + \pi r_0^2 \big(U(1) - U (0) \big)\\
& \equiv 2\pi r_0 \gamma - \pi r_0^2 | \Delta U |
\end{split}
\end{equation}
where we have assumed a sharp interface between saturated gene drive and wild-type states.
Here, $\Delta U$, the ``energy'' difference between the gene drive and wild type, causes the droplet to expand, and the role of an energy barrier to nucleation is played by the line tension term $\gamma$ \cite{barton1985analysis}.
This is indeed the case. For simplicity, we illustrate the nucleation approach with the cubic reaction term given by Eq.~7 in the main text. 

\begin{figure}[t!]
\centering
\includegraphics[clip,width=0.5\columnwidth]{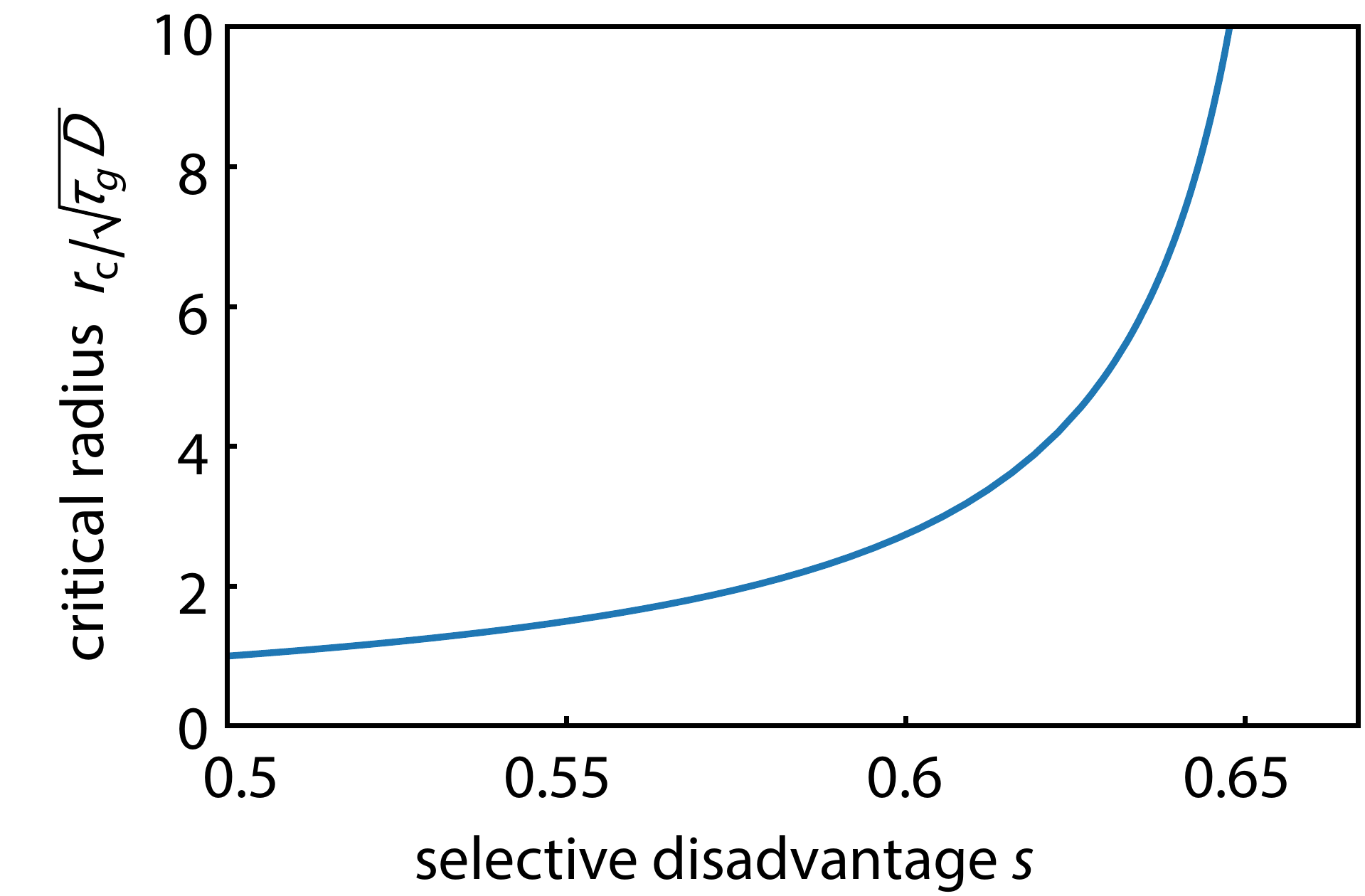}
\caption{
The critical radius of the nuclei $r_{\rm{c}}$ as a function of the selective disadvantage $s$.
}
\label{Fig_S4}
\end{figure}

First, we assume the logistic form of the spatial profile derived in the 1d limit by Barton and Turelli \cite{barton2011spatial}
\begin{equation}
q(r) = \frac{1}{1+e^{\sqrt{s/2 \tau_g D}(r-r_0)}},
\end{equation}
and the line tension term is
\begin{equation}
\gamma =  \int_0^{\infty} dr \frac{D}{2} \Big( \frac{dq}{dr} \Big)^2
= \frac{\sqrt{sD/2\tau_g } (e^{3 r_0 \sqrt{s/2 \tau_g D}} + 3 e^{2 r_0 \sqrt{s/2 \tau_g D}} ) }{12(e^{r_0 \sqrt{s/2 \tau_g D}}+1)^3}
\approx \frac{\sqrt{sD /2 \tau_g }}{12},
\end{equation}
in the limit of $1 \ll r_0 \sqrt{s/2\tau_g D}$.
The energy difference is given by
\begin{equation}
\Delta U = U(1) - U(0) = \frac{3s-2}{12 \tau_g},
\end{equation}
and the critical radius of the nucleus $r_{\rm{c}}$ which corresponds to the saddle point barrier of the free energy landscape is
\begin{equation}
r_{\rm{c}} = \frac{\sqrt{s \tau_g D/2}}{2-3s}
\end{equation}
as plotted in Fig.~\ref{Fig_S4}.
This result shows the divergence of $r_{\rm{c}}$ in the limit of $s\rightarrow s_{\rm{max}} (=2/3)$ and the above approximation ($r_0 \sqrt{s/ 2 \tau_g D} \gg 1$) becomes exact in this limit.
The diverging $r_{\rm{c}} (s)$ shown in Fig.~\ref{Fig_S4} is qualitatively consistent with the behavior found for the simplified gene drive initial condition in two dimensions shown in Fig.~\ref{Fig_S3} in the limit $q_0 \rightarrow 1$

\FloatBarrier\subsection{Wave velocities of the excitable waves}

\begin{figure}[htp]
\centering
\includegraphics[clip,width=0.5\columnwidth]{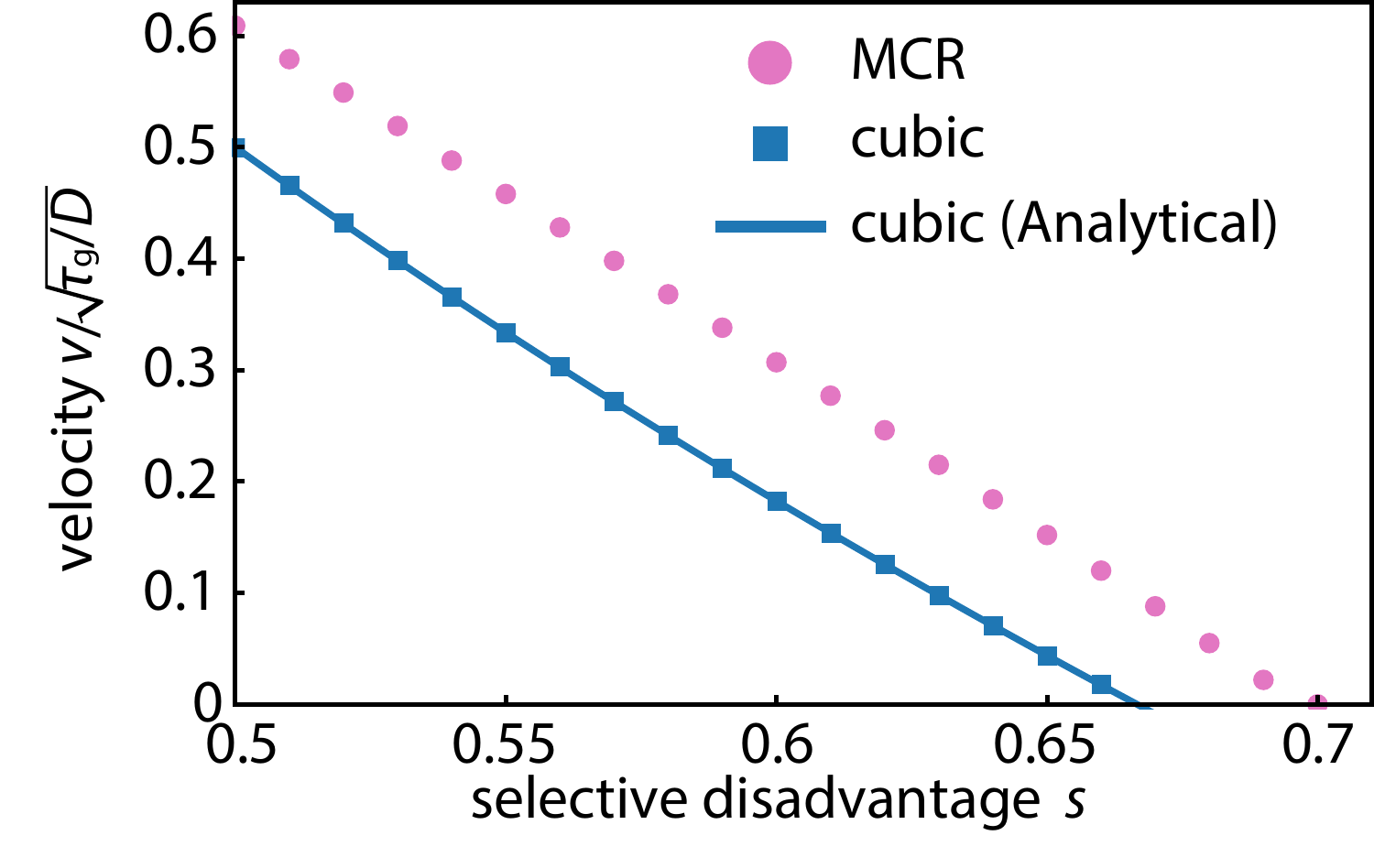}

\caption{
Asymptotic wave velocities $v$ of the excitable waves are plotted as a function of selective disadvantage $s$. 
The pink circular dots are numerically calculated wave velocities for the MCR model.
The blue curve is an analytically derived result for the simple cubic approximation, $ v(s)=(2-3s) \sqrt{D/2\tau_g s} $ \cite{barton2011spatial} and the blue squares are from numerical calculations, which confirm good agreement with the analytical result.}
\label{Fig_S5}
\end{figure} 

The reaction-diffusion equation admits traveling wave solutions with a continuous family of velocities. It selects the slowest speed asymptotically in the large time limit \cite{van2003front}.
The pink circular dots in Fig.~\ref{Fig_S5} are numerically calculated asymptotic wave velocities for the MCR model in the pushed wave regime.
We also plot the known wave velocity for the cubic approximation $ v(s)=(2-3s) \sqrt{D/2\tau_g s} $ \cite{barton1979dynamics, barton1989adaptation, barton2011spatial} for comparison.
Due to the larger reaction term $f_{\rm{MCR}}(q)>f_{\rm{cubic}}(q)$ (see discussion in Fig.~5), the wave velocity for the MCR model is always faster than the cubic approximation given the same selective disadvantage $s$.
In both cases, a larger selective disadvantage $s$ decreases the wave velocity, which eventually becomes zero at $s_{\rm{max}} = 0.697$ for the MCR model and the slightly smaller value $s_{\rm{max}}=2/3$ within the cubic approximation. \\

\FloatBarrier
\subsection{Calculation of the speed of the excitable waves}
In this section, we review the numerical method for calculating the speed of the excitable waves, following \cite{barton2011spatial}.
First, we assume a traveling waveform of the solution
\begin{equation}
q(x,t)=Q(x-vt)=Q(z),~z\equiv x-vt,
\end{equation}
with boundary conditions
\begin{equation}
\begin{split}
Q(z)\rightarrow 1~(z \rightarrow - \infty),~Q(z)\rightarrow 0~(z \rightarrow + \infty),\\
\frac{dQ}{dz} \rightarrow0~(z \rightarrow \pm \infty).
\end{split}
\end{equation}
By substituting $Q(z)$ into
\begin{equation}
\tau_g \frac{\partial q}{\partial t} = \tau_g D \frac{\partial^2 q}{\partial x^2} + R[q],
\end{equation}
we obtain
\begin{equation}
0=\tau_g D \frac{d^2 Q}{d z^2} + v \tau_g \frac{d Q}{d z} + R[Q].
\end{equation}
If we define the gradient $G$ as a function of $Q$, $G[Q] \equiv \frac{dQ}{dz}$ we arrive an ordinary differential equation
\begin{equation}
0=\tau_g D G\frac{dG}{dQ} + v \tau_g G + R[Q],
\end{equation}
with boundary conditions
\begin{equation}
G[0]=G[1]=0.
\end{equation}
It is known that there exists a unique velocity of the excitable wave $v$ that has solution $G[Q]$ of the above differential equation with the boundary condition \cite{keener1998mathematical}. 
We used a shooting method to determine such $v$ and plotted the results in Fig.~\ref{Fig_S5}.

\FloatBarrier
\subsection{Critical barrier strength}
\begin{figure}[b!]
\centering
\includegraphics[clip,width=0.5\columnwidth]{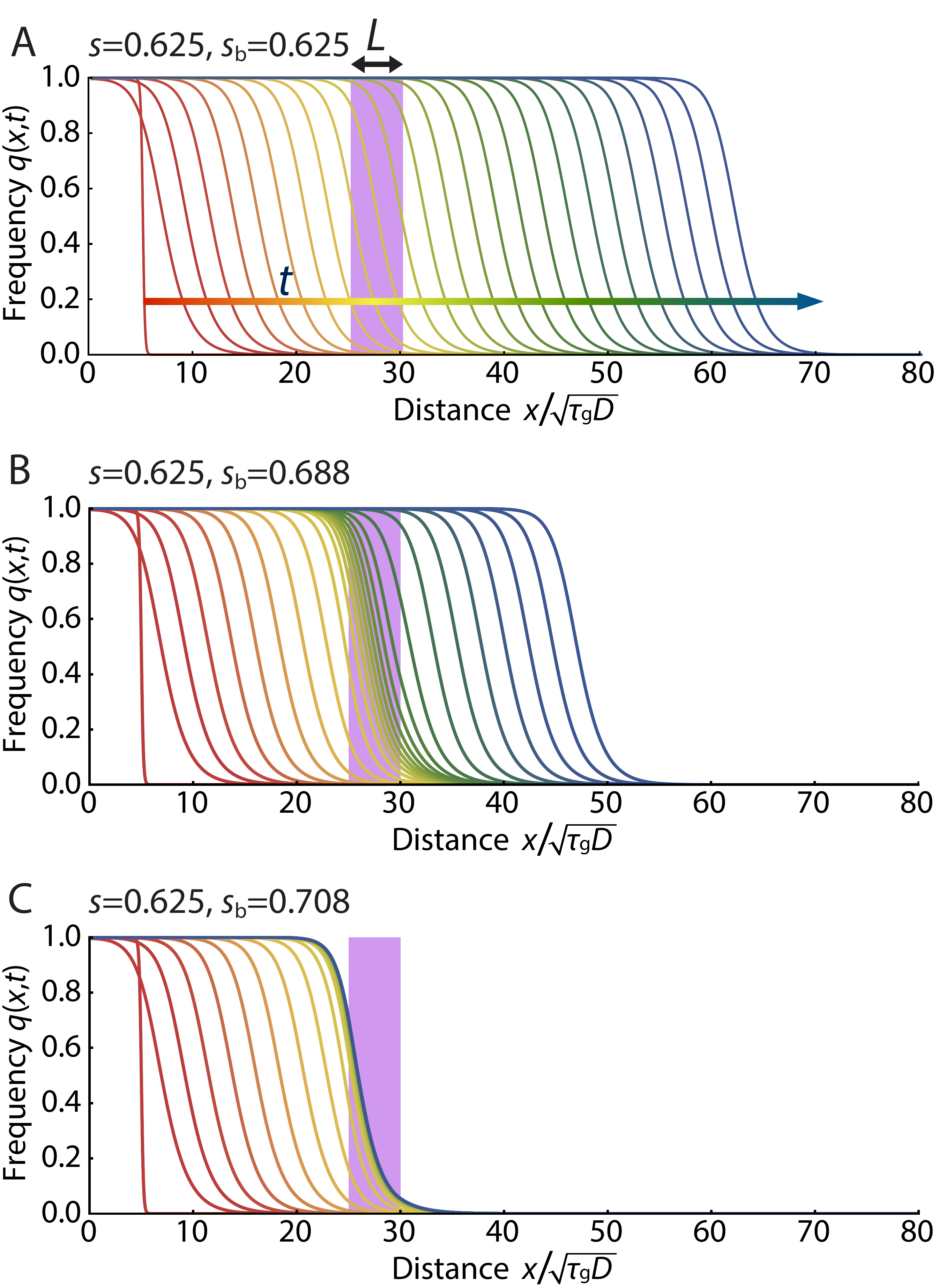}
  
\caption{
Stopping power of a selective advantage barrier in one dimension.
Numerical solutions of Eq.~5 are shown with time increment $\Delta t = 10.0 \tau_g$. 
The early time response is shown in red with later times in blue.
The selective disadvantage of the barrier is $s_b$ within the purple bar of width $L=5$ occupying the spatial region $25\sqrt{\tau_g D}<x<30\sqrt{\tau_g D}$ (shaded in blue) and $s=0.625$ otherwise. 
(A) The excitable wave propagates with constant speed when the barrier vanishes for $s_b = 0.625$.
(B) With $s_b =0.688>s=0.625$,
 the wave significantly slows down at the barrier, but recovers and propagates onwards. 
(C) The excitable wave is stopped when $s_b =0.708$.}
\label{Fig_S6}
\end{figure}
Fig.~\ref{Fig_S6} shows how the excitable wave can be slowed down and finally stopped by increasing the strength of a selective disadvantage barrier $s_b > s$.
As a reference, we first show dynamics of the excitable wave without a barrier ($s_b = 0.625$ matches the selective disadvantage $s=0.625$ outside) in Fig.~\ref{Fig_S6}\emph{A}.
When a small barrier is erected ($s_b=0.688<0.697$), the excitable wave significantly slows down within the barrier as expected from the results shown in Fig.~\ref{Fig_S5}. However, the wave recovers and propagates through the barrier as in Fig.~\ref{Fig_S6}\emph{B}.
When the barrier strength exceeds a critical value (in Fig.~\ref{Fig_S6}\emph{C} we plot the case $s_b = 0.708$) the excitable wave is stopped.

\begin{figure}[t!]
\centering
\includegraphics[clip,width=0.5\columnwidth]{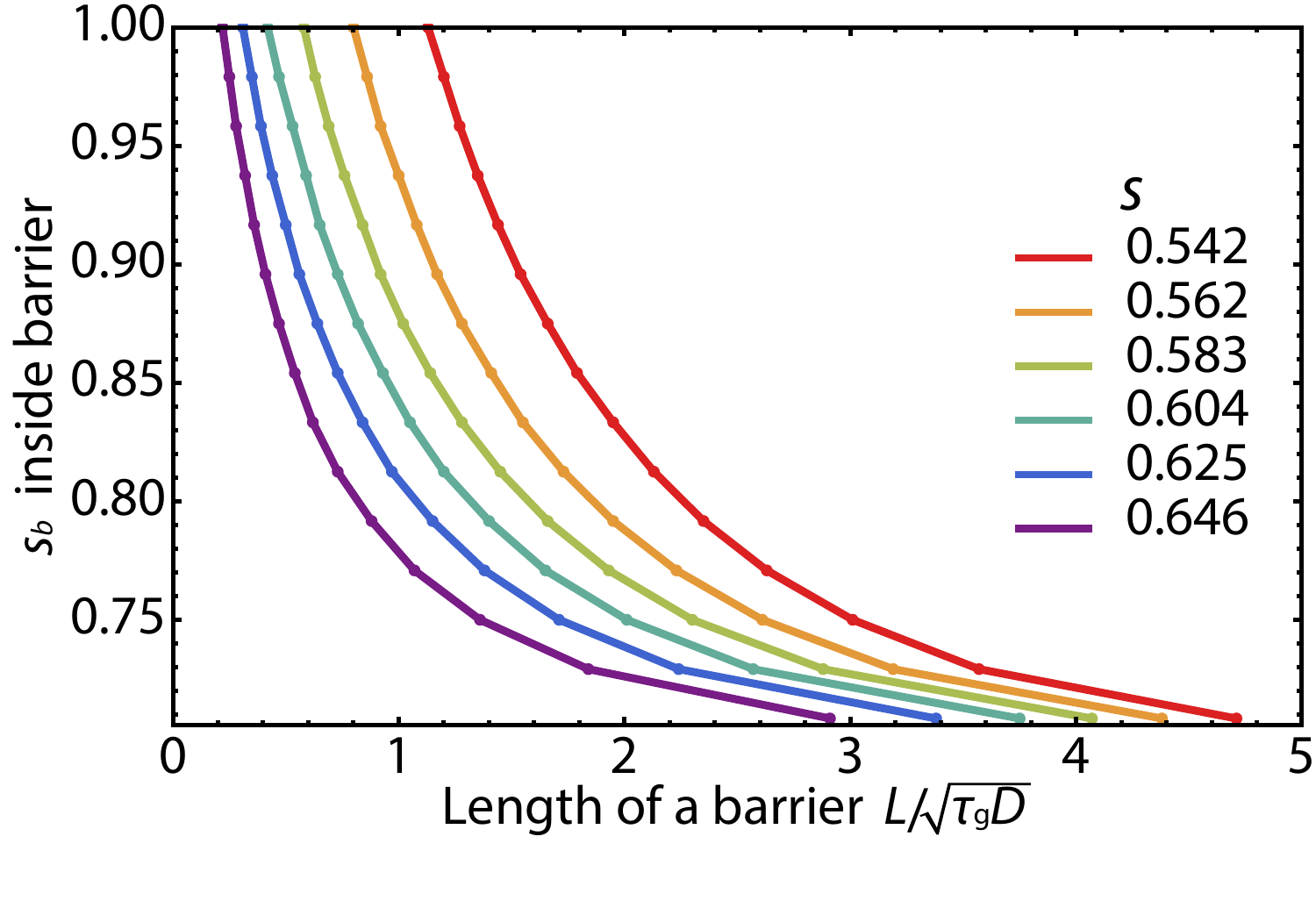}
\caption{
Critical width $L$ and the selective disadvantage $s_b$ of a barrier that is just sufficient to stop a pushed gene drive wave in one dimension.
The values are numerically obtained by placing the barrier in a region $25\sqrt{\tau_g D}<x<(25+L)\sqrt{\tau_g D}$.
Results are plotted for a variety of selective disadvantages $s$ outside the barrier region.
Given $s$, the excitable population wave can be stopped by a barrier whose parameters $(s_b, L / \sqrt{\tau_g D})$ lie above the curves.
}
\label{Fig_S7}
\end{figure} 
In Fig.~\ref{Fig_S7}, we plot the critical width $L$ and selective disadvantage within the one dimensional barrier region $s_b$ just sufficient to stop the excitable population wave of the gene drive species.
The values are numerically obtained by placing the barrier in a region $25\sqrt{\tau_g D}<x<(25+L)\sqrt{\tau_g D}$.
For example, when the selective disadvantage outside the barrier region is set to be $s \approx 0.65$, the excitable gene drive wave can be stopped by increasing $s$ by $\sim 20\%$ within the barrier region of thickness $\sim \sqrt{\tau_g D/s}$.

\FloatBarrier
\subsection{Fluctuations due to finite population size}
In this section, we estimate effects of fluctuations due to a finite population size using mosquitos as an example.
First, we define the effective spatial population size $N_{\rm{eff}}$ to be the number of mosquitos with which an individual might conceivably mate during its generation time $\tau_g$ \cite{hartl1997principles}.
Given a diffusion constant $D$, the two dimensional area an individual can explore during its life time $\tau_g$ is $\pi (\sqrt{4D\tau_g})^2$ and the effective population size in two dimensions is 
\begin{equation}
N_{\rm{eff}} \equiv 4 \pi D \tau_g n,
\end{equation}
where $n$ is the area density of organisms.
Here, we estimate $N_{\rm{eff}}$ using parameters appropriate to mosquitos: $\tau_g\sim 10[\text{days}]$ \cite{deredec2011requirements}, $D\sim 0.1 [\text{km}^2/\text{day}]$ and $n\sim 1 [\rm{m}^{-2}] = 10^6 [\rm{km}^{-2}]$ to get $N_{\rm{eff}}\sim 10^5 - 10^6$.
With such a large effective population size, we believe that the dynamics can be well described by the deterministic limit explored here.
Fluctuations \emph{can} play a role for systems with smaller populations and such effects have been thoroughly investigated in the physics literatures \cite{brunet1997shift, van2003front, brunet2015exactly, cohen2005fluctuation, hallatschek2009fisher}.
Pulled waves are more sensitive to fluctuations, with a Fisher wave velocity that changes according to  
\begin{equation}
v=v_F[ 1 - O(1/\ln^2 N_{\rm{eff}})],
\end{equation}
where $v_F$ is the velocity of the pulled wave in the deterministic limit \cite{brunet1997shift}.

\clearpage
\twocolumngrid
\bibliography{reference} 
\end{document}